# A Hybrid Quantum Mechanics Machine Learning Forcefield (QM/ML) Framework for Accurate Solute–Dislocation Interaction Simulations

Junting Zhang[1], Colleen Reynolds[1], Ryan S. Stroud[1], Jana Smutna[1], Daniel J. M. King[1], Andrew P. Horsfield[2] and Mark R. Wenman[1]

1. Department of Materials and Centre for Nuclear Engineering, Imperial College London, London, UK
2. Department of Materials and Thomas Young Centre, Imperial College London, London, UK

Corresponding author: Mark R. Wenman
Corresponding author email: m.wenman@imperial.ac.uk

## Abstract:

Solute–dislocation interactions play a central role in controlling microstructural evolution and mechanical behaviour of structural materials, yet conventional atomistic modelling approaches struggle to combine the chemical accuracy with computational scalability. In the nuclear industry, these challenges become particularly acute, as experiments reveal strong correlations between solute segregation and irradiation-induced dislocation loops. However, theoretical insight remains limited because density functional theory (DFT) simulations are prohibitively expensive at relevant length scales, while traditional semi-empirical interatomic potentials lack the chemical fidelity required for predictive solute–defect calculations. Here, we introduce a hybrid quantum-mechanics/machine-learning (QM/ML) simulation framework that couples DFT with neural-network machine learning interatomic potentials (MLIPs), enabling accurate atomistic dislocation simulations at reduced computational cost. We demonstrate the QM/ML framework's capability by reproducing the experimentally observed Sn and Fe segregation to dislocation loops in Zr and investigating magnetically complex solute-dislocation interactions in steel. These results establish the approach as a transferable, high-fidelity tool for modelling irradiation-induced defect structures and benchmarking emerging MLIPs.

# Introduction

Interactions between solute atoms and dislocations govern microstructural evolution and mechanical degradation across structural alloys [1-3]. Segregation of solutes to dislocation lines and loops can pin or drag dislocations and modify glide, cross slip, and climb [1, 4, 5]. Solute decorated cores can also nucleate precipitates or secondary phases [6-9]. Together these processes control yield strength, creep resistance, swelling, ductility, and ultimately determine service lifetime [10, 11]. An accurate description of solute dislocation interactions is therefore essential for predicting long term mechanical performance under demanding service conditions.

In engineering alloys, dislocations rarely exist in isolation but evolve within complex defect and chemical environments that depend on processing and service conditions [12-13]. This coupling is particularly pronounced under irradiation, where continuous generation of point defects leads to the formation of high densities of dislocation loops that act as sinks for solutes and impurities [14-18]. Experimental studies in zirconium cladding alloys [19,20] and reduced activation ferritic martensitic (RAFM) components [21-23] have revealed pronounced solute segregation and clustering at irradiation-induced loops, linking defect evolution to mechanical degradation [24-30]. However, while such observations provide detailed structural and chemical snapshots, they do not directly resolve the mechanisms governing solute binding and migration, motivating the need for predictive atomistic modelling.

Accurate predictions of solute–dislocation interactions require treatment of two regimes. First, short-range core chemistry demands a description of charge localisation, bonding and magnetic effects [4,31,32]. Second, the long-range elastic field of a dislocation requires simulation cells containing thousands of atoms to be represented faithfully [4,33,34]. Density functional theory (DFT) provides the most reliable description of core chemistry, capturing electronic and magnetic contributions to solute binding and local stacking-fault configurations [4,32,34–37]. However, its computational cost prevents routine simulations of cells large enough to include realistic dislocation elasticity; even earlier studies with >1400 atoms remain costly and difficult to reproduce, and probably still have strain effects from the boundary conditions [4,32,34,37]. At the opposite extreme, empirical interatomic potentials such as embedded-atom-method (EAM) models enable molecular dynamics (MD) simulations containing millions of atoms and can reproduce qualitative loop evolution under thermal and irradiation conditions [33,38–41]. Yet they lack the electronic and magnetic fidelity required for accurate defect chemistry, particularly in BCC Fe, and they cannot be applied to systems containing elements that were not included in their original fitting [42].

Machine-learning interatomic potentials (MLIPs) provide a promising alternative for dislocation simulations. Gaussian approximation potentials (GAP), moment tensor potentials (MTP), Atomic Cluster Expansion (ACE) and the latest graph-neural-network (GNN) models trained on large DFT datasets have achieved near first-principles accuracy at computational costs comparable to empirical potentials, and have resolved long-standing discrepancies in dislocation core structures and Peierls barriers [43–49]. Yet their accuracy remains strongly dependent on the availability of relevant training data. Universal pretrained models such as MACE now span more than eighty elements, but the underlying datasets rarely include defects or solute-decorated dislocation configurations, limiting their reliability at complex defect cores [50]. As a result, the accuracy of MLIPs for solute–dislocation chemistry often remains uncertain, and fine tuning for specific problems is difficult to validate because corresponding DFT reference calculations are prohibitively expensive [51,52].

Hybrid quantum-mechanics/molecular-mechanics (QM/MM) schemes provide one route to combine the accuracy of DFT with the scalability of empirical potentials [53–55]. Studies on W and Fe dislocations have demonstrated the capacity of such methods to describe solute trapping, diffusion and Peierls stresses using several hundred DFT atoms embedded within thousands of molecular-mechanics (MM) atoms [53–55]. However, traditional QM/MM approaches inherit limitations from both components. The DFT region remains computationally intensive, especially for magnetic systems, and the accuracy of the classical region typically restricts chemical fidelity. Mixing energies and forces between QM and MM regions requires close compatibility between the two descriptions, which is rarely achieved with empirical potentials and often leads to significant errors [42,56,57]. These challenges have motivated efforts to replace classical potentials with early stage MLIPs in hybrid frameworks [58,59]. Earlier QM/ML implementations were based on linear MLIP formulations, which generally required retraining for each new defect environment and were typically applied only to a limited set of materials, restricting the overall transferability of these approaches [58,59]. More recent hybrid ML-based approaches, including ML/MM treatments of molecular environments and spatial mixing of MLIPs for localised defects, have shown that locally adaptive descriptions can improve both efficiency and accuracy [75, 76]. However, these methods address different regimes from chemically sensitive solute interactions with extended crystalline defects, where an explicit quantum-mechanical description remains essential to capture local bonding, charge redistribution and magnetic effects, while preserving the long-range elastic response of the surrounding crystal.

In this work we introduce a new hybrid QM/ML approach designed specifically for solute–dislocation interaction studies, using the more recent universal NN based MLIPs. The framework combines DFT-level accuracy for defect-core chemistry with the long-range elasticity and transferability of a modern MLIP, without requiring retraining for each system. This method fills the gap among conventional modelling methods by providing chemically consistent forces across the QM and ML regions and enabling simulations large enough to capture realistic dislocation geometries. To benchmark the method and demonstrate its capability, we apply it to solute–dislocation binding in Zr and Fe, which provide experimentally grounded and scientifically demanding test cases. The approach also enables benchmarking of emerging MLIPs in regimes where high-quality DFT data are difficult to obtain, offering a route to validate MLIP behaviour for complex irradiation-induced defect structures.

# Results

## Pre QM/ML model convergence test:

The QM/ML framework requires convergence testing of both conventional DFT parameters and region-specific parameters, in particular the sizes of Zones I and II, to ensure that defect-core chemistry is accurately captured. As the method relies on force-based relaxation, convergence was assessed using the maximum absolute force and the corresponding force-error trends.

A dislocation disk model with a radius of 100 Å containing an ⟨a⟩ type screw dislocation in hcp Zr was first pre-relaxed using the MLIP until the maximum force component fell below 0.1 eV $Å^{-1}$. A central region with a radius of 20 Å was then extracted for DFT evaluation, and the resulting atomic forces were analysed as a function of distance from the dislocation line (Fig. 1a). Although the MLIP-relaxed structure satisfies the force convergence criterion, DFT reveals significant discrepancies in force predictions in regions of structural complexity, particularly within the dislocation core (<7 Å) and near the free surface (>14 Å), where vacuum-induced boundary effects become significant. In contrast, good agreement between MLIP and DFT forces was observed in the intermediate region between 7 and 14 Å. Based on this analysis, a QM region radius of 10 Å was selected for subsequent simulations.

To ensure that forces within Zone I are not affected by the artificial surface, the thickness of the Zone II inner buffer was systematically varied from 4 to 14 Å with a fixed Zone I radius of 10 Å. Force errors for atoms in Zone I were evaluated relative to a 14 Å reference case (Fig. 1b). When the buffer thickness exceeded 7 Å, the maximum force error within Zone I decreased below 0.05 eV $Å^{-1}$. An inner buffer thickness of 8 Å was therefore adopted as a balance between accuracy and computational cost. These tests define the DFT zones size required for reliable energy and force evaluation, and highlight two primary sources of error: inaccuracies in MLIP force predictions in dislocation core regions, and artificial forces arising from vacuum boundaries.

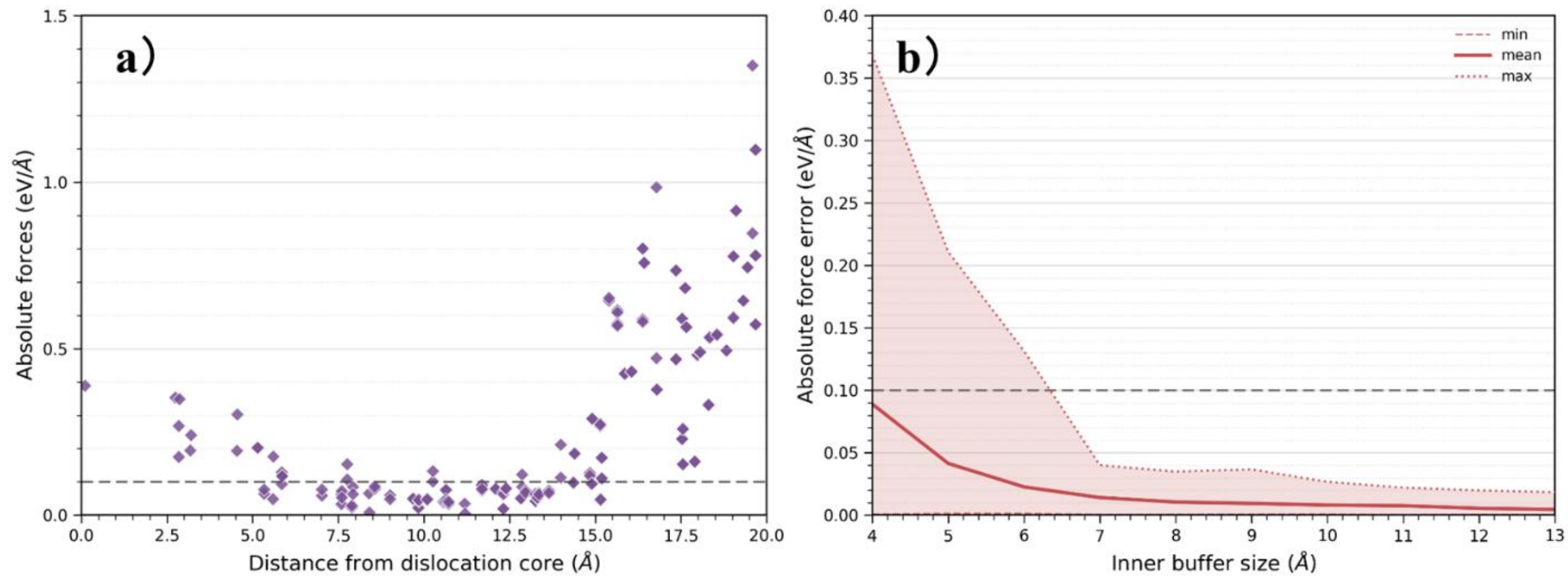

*Figure 1: a) DFT absolute forces as a function of radial distance, for a MLIP relaxed model. The dashed line denotes a maximum force components of 0.1 eV $Å^{-1}$ predicted by MLIP. b) DFT absolute force error for Zone I atoms as a function of Zone II inner buffer thickness, with a 10 Å Zone I radius. Forces are referenced to the 14 Å Zone II case.*

In addition to the vacuum surrounding the DFT zones, a larger vacuum region is present beyond Zone IV. While flexible boundary conditions based on lattice Green's functions have been proposed in previous QM/MM studies to mitigate such effects, their application to systems containing dislocations and stacking faults is non-trivial and limits transferability [55]. Instead, a fixed-boundary approach was adopted, in which atoms within a specified distance from the outer boundary are constrained during relaxation. Although residual boundary and vacuum-induced errors cannot be completely eliminated, the quantities of interest in this work are energy differences, in particular solute–dislocation binding energies. These are evaluated between configurations with closely related geometries and boundary conditions, such that systematic errors arising from the vacuum and constrained regions are expected to largely cancel.

The effect of this boundary treatment was assessed using MLIP-only calculations of solute–dislocation binding energies in α-Zr (Fig. 2). Binding energies for Fe interstitials and Sn substitutional solutes interacting with screw and edge dislocations were evaluated as a function of the radius of unconstrained atoms. When too few atoms were fixed, noticeable fluctuations in binding energy were observed. A plateau region was reached as the number of constrained boundary atoms increased, indicating stable and converged behaviour. However, when all atoms were allowed to relax, the presence of high-energy, poorly coordinated boundary atoms led to significant deviations in the calculated binding energies. These results demonstrate that appropriate boundary constraints are essential for obtaining reliable energetics. In this work, atoms within 30 Å of the outer boundary were fixed during QM/ML production runs to ensure stability and accuracy.

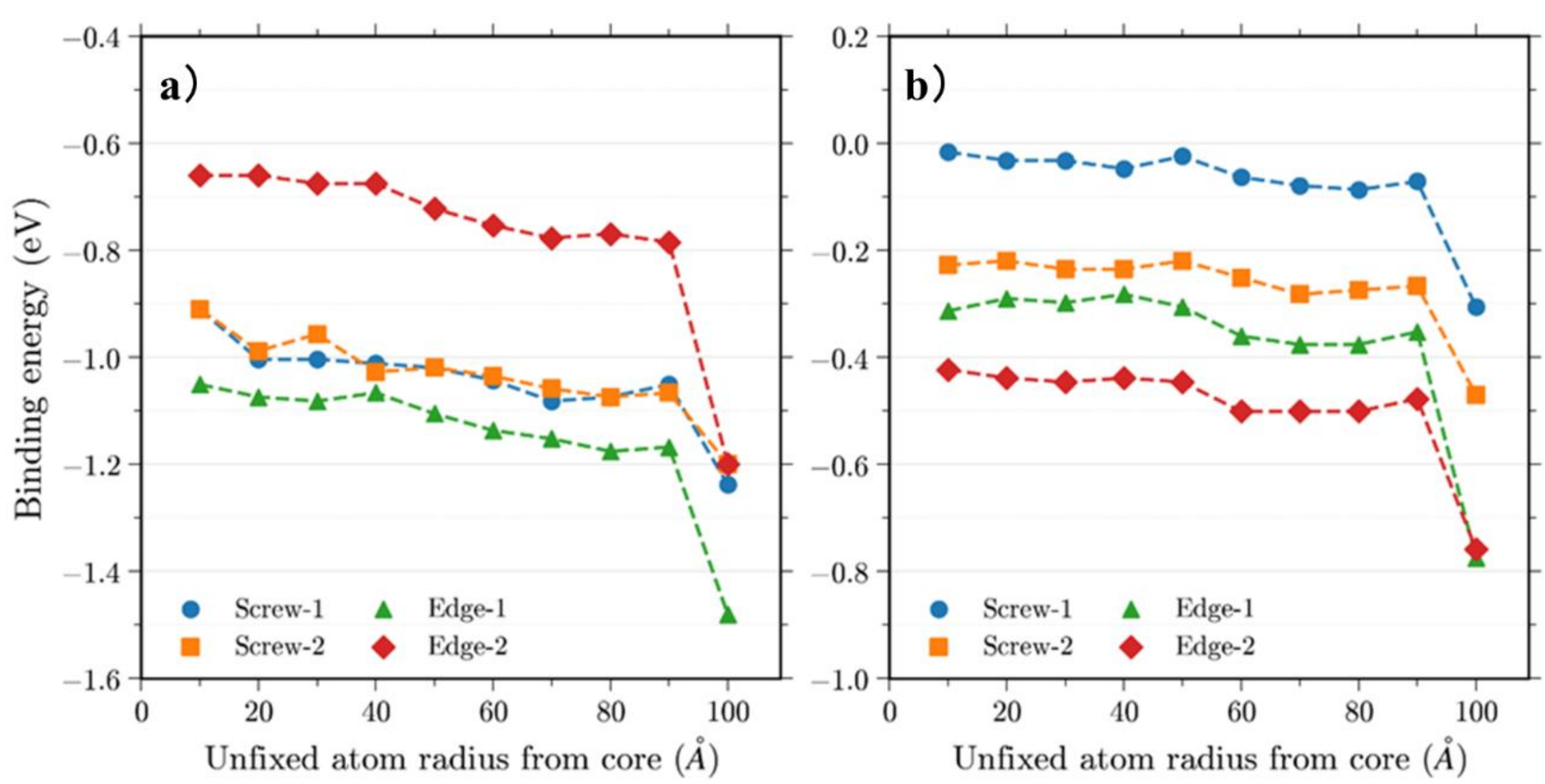

*Figure 2: Fixed Zone IV outer buffer convergence test. Solute–dislocation binding energies calculated using MLIP-only simulations for (a) Fe and (b) Sn solutes interacting with screw and edge dislocations in* α*-Zr. In each case, two solutes are bound to the same dislocation. Binding energies are plotted as a function of the radius of unconstrained atoms, varied from 10 to 100 Å within a dislocation disk of radius 100 Å, corresponding to progressively reducing the number of fixed boundary atoms.*

Separate MLIP benchmarks were carried out for elastic constants, point-defect formation energies and lattice parameters. The underlying potentials reproduce acceptable elastic behaviour and defect energetics, indicating that they are suitable for describing the dislocation strain field and solute chemistry in the perfect lattice. Detailed numerical values are provided in the Supplementary Information. The fine-tuned neuroevolution potential (NEP-ft) model was designed to improve on the universal NEP potential, which shows less reliable predictions than the universal (MACE potential. In the NEP-ft potential, the elastic properties are improved to an acceptable range. For defect formation energies, MACE and NEP-ft show good agreement with DFT in Zr, and for substitutional solutes and vacancies in BCC Fe. However, for C interstitials in Fe, the formation energy error reaches about 0.3 eV for MACE and up to 0.8 eV for NEP-ft.

Although the NEP-ft potential performs worse for BCC Fe elastic properties and defect energies overall, it exhibits one distinct advantage in its lattice-parameter prediction. As shown in Table 1, the universal MACE potential provides excellent agreement with DFT lattice parameters for Zr, but it overestimates the BCC Fe lattice parameter by up to 1.6 %. This overestimation introduces a significant artificial strain when used in hybrid QM/ML dislocation simulations, which in turn affects the energies of large defects such as dislocations.

*Table 1: Lattice parameter predictions from the universal MACE and NEP-ft potentials for Zr and Fe, compared with DFT.*

| Method | Material | a lattice parameter Å | a lattice parameter error | c lattice parameter Å | c lattice parameter error |
|---|---|---|---|---|---|
| **DFT** | **Zr** | 3.23 | -------- | 5.17 | -------- |
| MACE | Zr | 3.23 | 0.03% | 5.19 | 0.33% |
| **DFT** | **Fe** | 2.83 | -------- | -------- | -------- |
| MACE | Fe | 2.88 | 1.59% | -------- | -------- |
| NEP-ft | Fe | 2.83 | -0.21% | -------- | -------- |

## QM/ML usage in Zr system

QM/ML solute–dislocation interactions were calculated for two Zr dislocation segments, a screw segment in a 1st order prismatic <a> loop and an edge segment containing a stacking fault representative of basal-plane loops. As shown in Fig. 3, the screw segment exists in a perfect <a> loop on the $(01\bar{1}0)$ plane with a Burgers vector of $\frac{1}{3}\langle 2\bar{1}\bar{1}0\rangle$, and the screw component therefore has both its line direction and Burgers vector aligned along $\langle 2\bar{1}\bar{1}0\rangle$. For the <c/2> loops on basal planes, which form under irradiation conditions, the accumulated vacancy plate introduces a high-energy stacking fault. Commonly, the metastable stacking fault containing loop tends to interact with partial dislocations, and undergo a stacking fault transformation to lower energy stacking faults like extrinsic or intrinsic stacking faults [33, 37]. In this work, the vacancy-type edge dislocation considered is one possible loop segment, with line direction $\langle 2\bar{1}\bar{1}0\rangle$ and a Burgers vector of $\frac{1}{2}[0001]$ in the extrinsic-fault configuration.

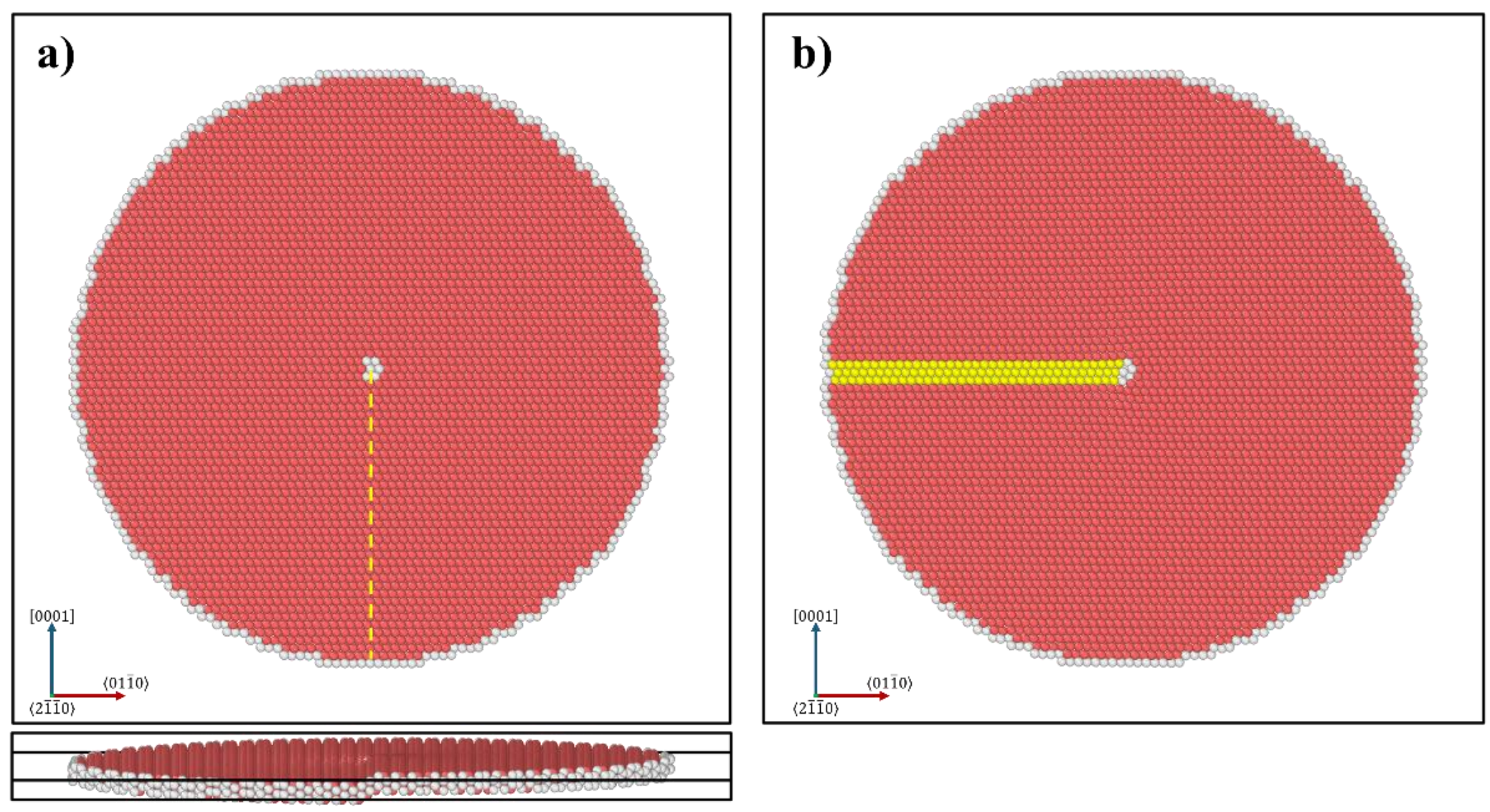

*Figure 3: Zr dislocation disks considered for sample calculations. a) a screw segment in the perfect first-prismatic <a> loop, and b) an edge segment in a <c/2> vacancy-type dislocation loop containing an extrinsic stacking fault, where the atoms in the faulted region are colored with yellow. The yellow dashed line denotes the slip plane of the screw dislocation.*

Solutes were inserted into MLIP-pre-relaxed dislocation structures. A random-sampling script was then used to generate 15 non-repeating substitutional or interstitial sites within a 10 Å radius of the dislocation core, corresponding to the Zone I QM region in the QM/ML calculation. This sampling provides broad coverage of the relevant local environments around the dislocation core while avoiding excessive redundancy from symmetry-related sites, and is sufficient to identify representative high- and low-energy configurations. At this stage, the MLIP is used primarily to sample the strain-defined distribution of candidate solute sites, while the subsequent QM/ML calculations restore the local chemistry. For each solute–dislocation system, the highest- and lowest-energy MLIP-predicted sites were then selected and carried forward to full QM/ML simulations for benchmarking.

Substitutional Sn and Nb and interstitial H and Fe binding energies for the two Zr dislocations are shown in Fig. 4. The overall range of binding energies for each solute is generally consistent across the two dislocation types. Fe solutes strongly bind to both dislocation cores, while Nb exhibits only slight attraction. Sn and H solutes show weaker binding and therefore more positive binding energies, indicating less favourable interaction and possible repulsion from the dislocation cores.

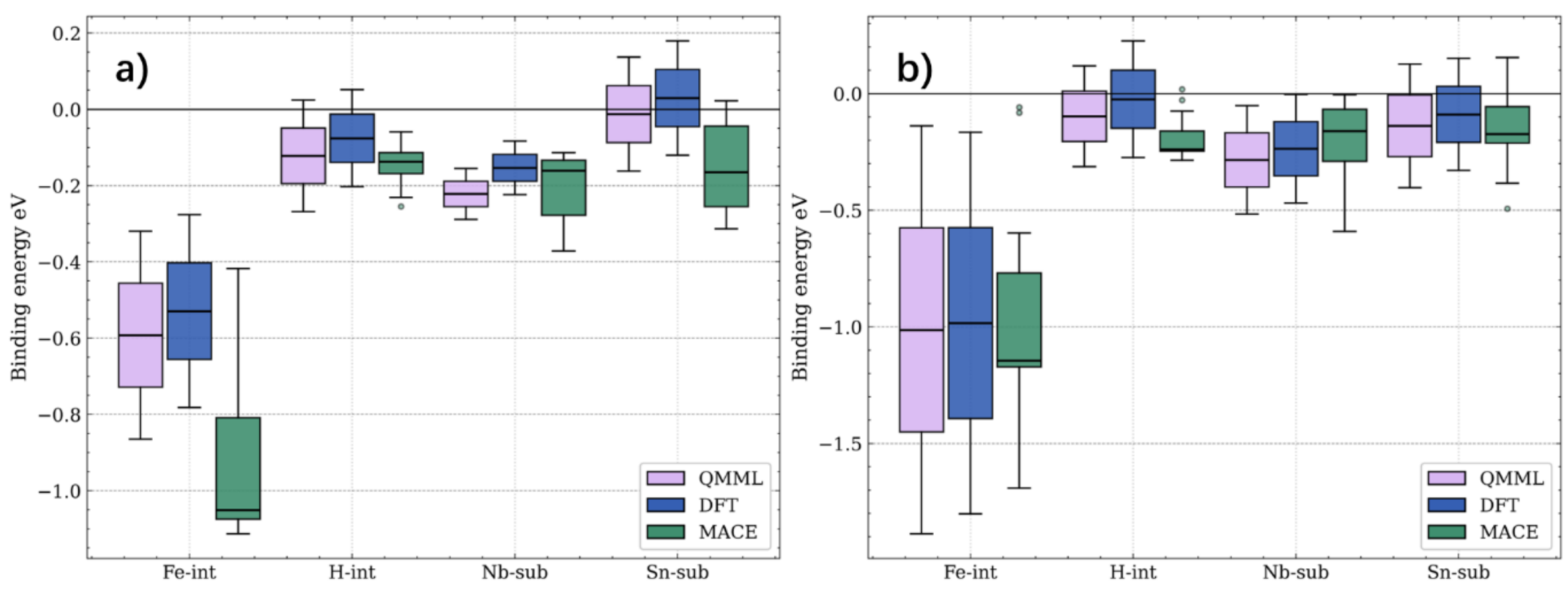

*Figure 4: Box plots of solute–dislocation binding energies for interstitial Fe and H, and substitutional Nb and Sn in Zr, interacting with (a) ⟨a⟩ type screw dislocation and (b) ⟨c/2⟩ vacancy-type edge dislocation containing an extrinsic stacking fault. For each solute–dislocation configuration, results are shown for the final QM/ML binding energies, the DFT component of the QM/ML energies, and the MLIP binding energies obtained after pre-relaxation.*

Previous DFT work showed that Fe interstitials in Zr lose their magnetic moment [31]. Since only Fe interstitials are considered here, the present DFT calculations were performed without spin polarisation. This approximation was checked by analysing the Fe-projected density of states near the Fermi level during the QM/ML calculations, providing a consistency check on the electronic-structure treatment and illustrating how QM/ML can recover information absent from MLIP-only simulations. Further details are provided in the Supplementary Information.

From a methodological perspective, comparison between QM/ML, DFT and MLIP binding energies helps distinguish local chemical errors from embedding effects introduced by the hybrid partition. Under the assumption that the dominant chemically sensitive contribution is captured within the DFT zones, the DFT results provide a reference for the local interaction, whereas the QM/ML values represent the corresponding embedded prediction including the long-range MLIP environment. The deviations of the QM/ML and MLIP binding energies from the DFT values, summarised in Fig. 5, are therefore not unambiguously errors: when the interaction is effectively localised within the DFT zones, they mainly reflect the additional elastic embedding absent from the finite DFT model; when the MLIP description is inaccurate, they instead reflect error in the MLIP-based description.

The QM/ML results show excellent agreement with DFT, with energy deviations typically below 0.1 eV. In contrast, the MACE-MLIP results exhibit significantly larger deviations. Although the MLIPs reproduce elastic properties and defect formation energies reasonably well in the bulk lattice, their accuracy deteriorates in the highly strained and chemically complex environment of dislocation cores, particularly for interstitial solutes. For example, in the case of Fe interstitials, the variation in the minimum binding energy for the same configuration can reach up to 0.3 eV.

More generally, MLIPs tend to overestimate solute–dislocation attraction, leading to excessively negative binding energies. This behaviour is associated with an underestimation of repulsive interactions at high-energy sites near the dislocation core. However, an exception is observed for Fe interstitials interacting with the ⟨c/2⟩ edge dislocation, where the MLIP underestimates the binding strength. This behaviour is likely related to known limitations of current MLIPs in describing stacking-fault energetics and their coupling to complex dislocation-core chemistry and interstitial-induced strain fields.

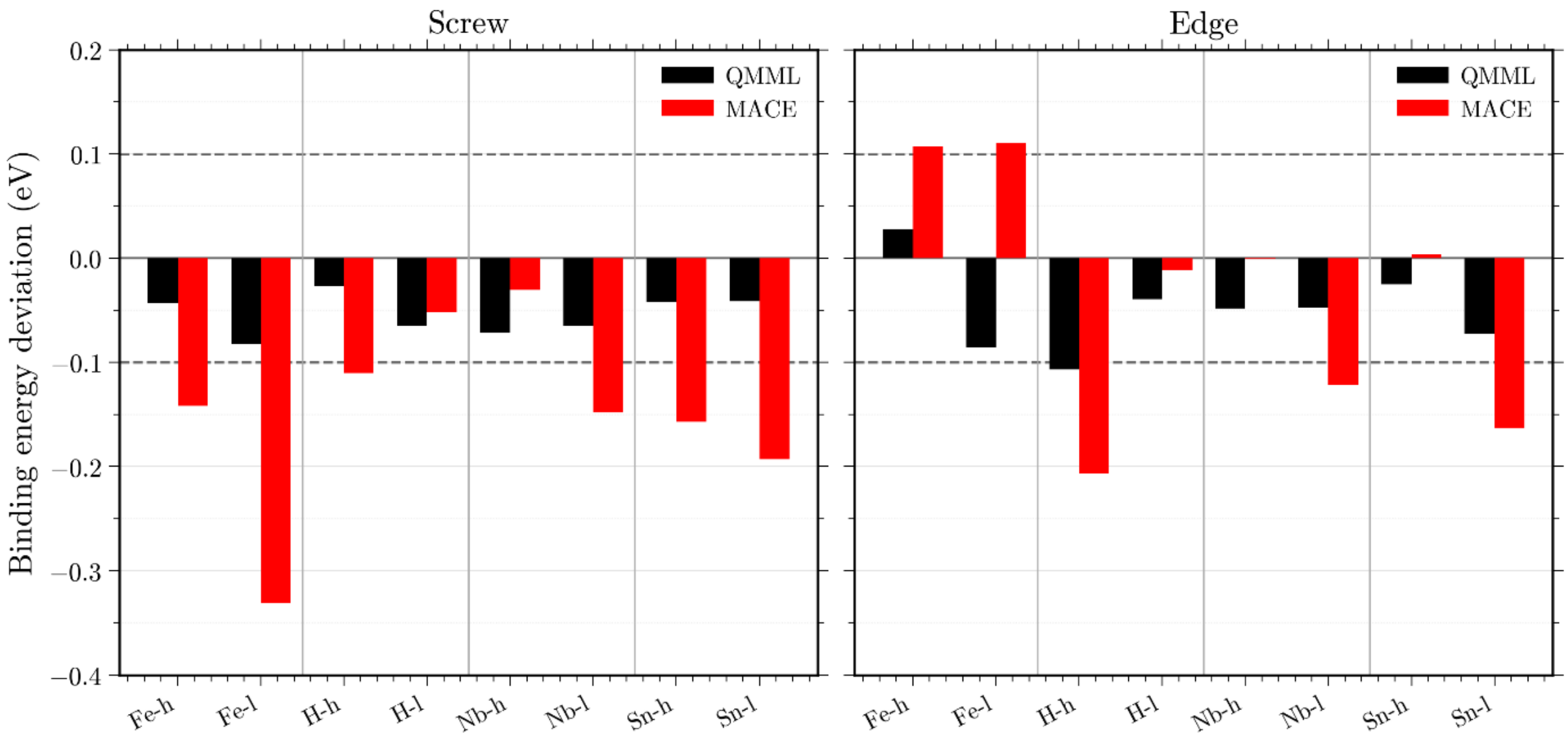

*Figure 5: Binding energy deviation for each solute–dislocation configuration in Zr, comparing QM/ML and MLIP-only results with DFT reference values. The screw dislocation values are on the left, and the edge dislocation values are on the right. "h" and "l" denote the high and low energy solute sites to the same dislocation.*

## QM/ML usage in magnetised BCC Fe system

Beyond the Zr systems, QM/ML simulations were also performed for the magnetised BCC Fe system. In addition to simple defect–dislocation interaction studies, NEP-ft potential was tested alongside the universal MACE potential to demonstrate the capability of the framework for benchmarking different MLIPs within a consistent QM/ML setting. For the Fe dislocations considered (Fig. 6), the supercell was oriented with x:[111], y:$[\bar{1}\bar{1}2]$ and z:$[1\bar{1}0]$. The screw and edge dislocation were then created with line directions [111] and $[1\bar{1}0]$, and Burgers vectors of $\frac{1}{2}[111]$.

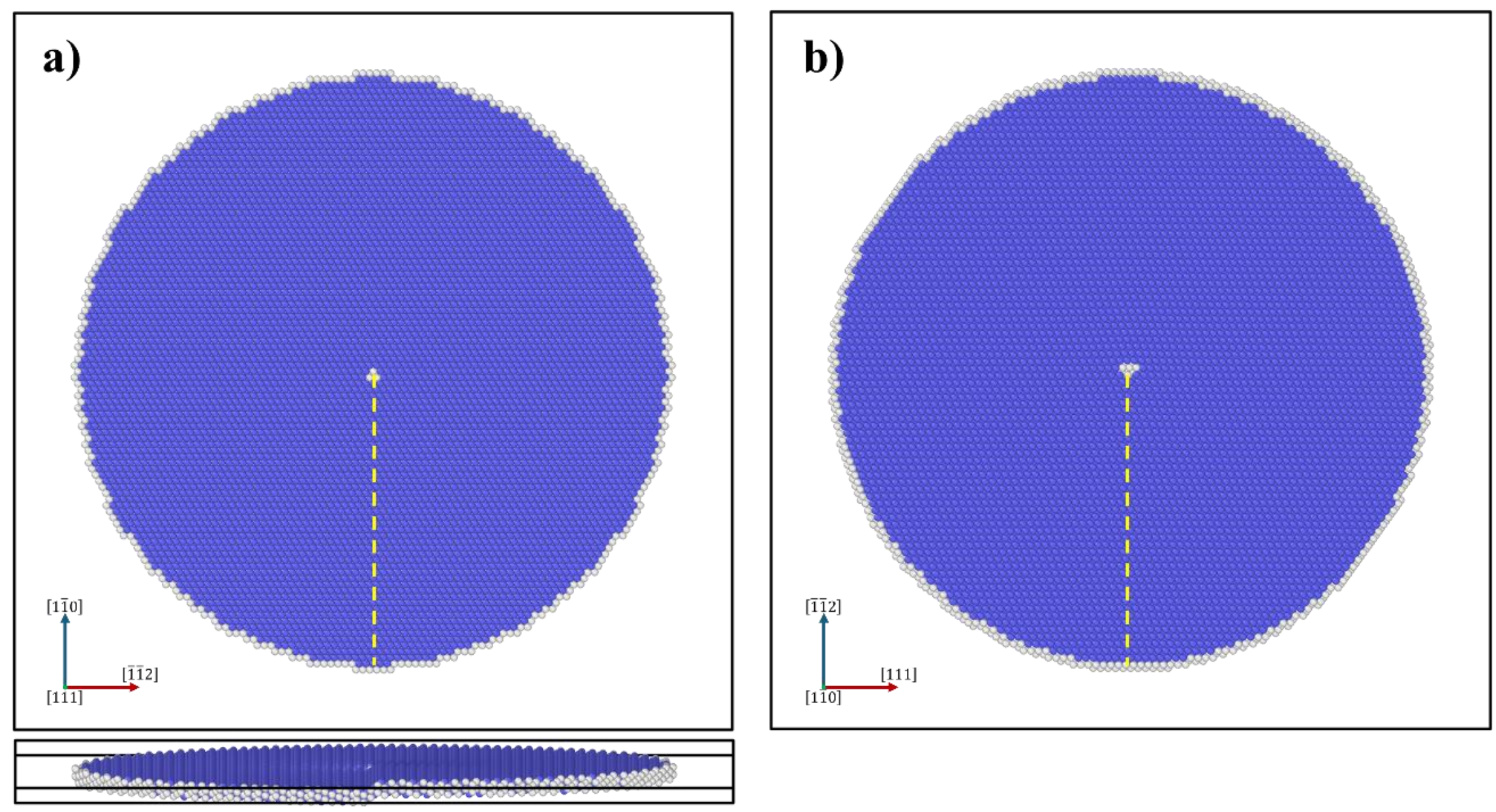

*Figure 6: BCC Fe dislocation configurations used in QM/ML simulations. a) screw and b) edge dislocation segments with Burgers vector 1/2[111] are shown. The yellow dashed line indicates the slip plane and the extra half-plane for the screw and edge dislocations, respectively.*

Fig. 7 shows the binding energies of substitutional Si and interstitial C to screw and edge dislocations in BCC Fe, calculated using the two MLIPs within the QM/ML framework. Consistent with the Zr results, interstitial solutes exhibit larger deviations in MLIP-only predictions, reflecting the limited representation of defect environments in the training data and the larger errors observed in interstitial defect formation energies. For substitutional Si, the QM/ML and DFT binding energies show good overall agreement, indicating weak attraction to both dislocation types. Differences between the two MLIPs remain relatively small, although the NEP-ft potential shows slightly larger deviations from the QM/ML reference.

In contrast, interstitial C exhibits a pronounced dependence on the choice of MLIP. The QM/ML binding energies obtained using the MACE and NEP-ft potentials differ significantly, with MACE predicting more negative (attractive) binding. This behaviour is directly linked to differences in lattice parameter predictions between the two potentials. The MACE potential overestimates the BCC Fe lattice parameter by approximately 2%, introducing an artificial tensile strain in the DFT zones. Unlike substitutional solutes, interstitial C generates a strong local strain field. This strain partially compensates the tensile mismatch introduced by the MLIP, stabilising the interstitial configuration and leading to more negative binding energies for otherwise identical geometries. In contrast, the improved lattice parameter agreement of the NEP-ft potential reduces this artificial strain effect, resulting in less negative and more physically consistent binding energies.

From a methodological perspective, these results highlight two important points. First, the QM/ML framework is sensitive to the underlying MLIP, particularly through its influence on lattice parameter and long-range strain. Second, the approach provides a practical means of diagnosing MLIP limitations in complex defect environments, where direct DFT benchmarking would otherwise be computationally prohibitive.

In addition to accuracy, the choice of MLIP also affects computational efficiency. The improved lattice-parameter agreement of the NEP-ft potential reduces the strain mismatch at the QM/ML interface, which may facilitate relaxation convergence. For the same structures, QM/ML simulations using NEP-ft require approximately 20% fewer total CPU hours than those using MACE. Detailed comparisons are provided in the Supplementary Information.

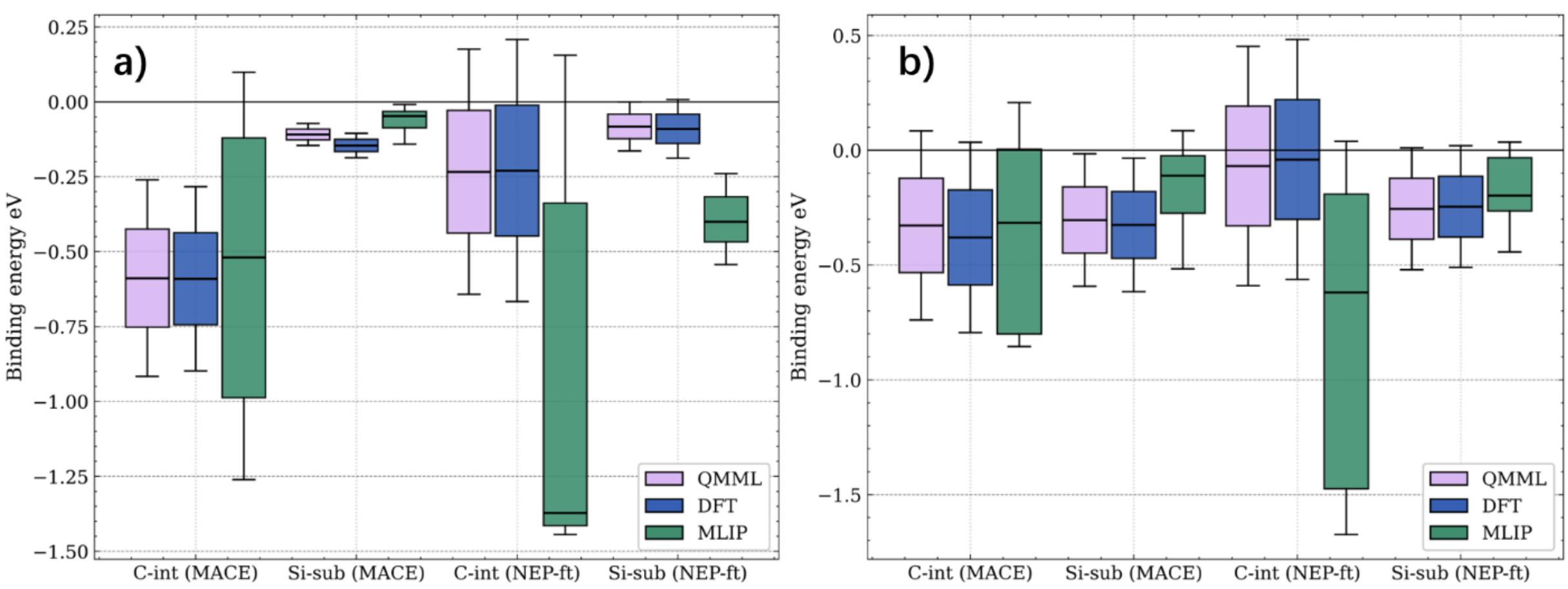

*Figure 7: Solute–dislocation binding energies from QM/ML production runs and MLIP pre-relaxations for interstitial C and substitutional Si interacting with (a) screw and (b) edge dislocations in BCC Fe. Results are shown for both the universal MACE and NEP-ft potentials.*

## Importance of model size stability

To benchmark MLIP accuracy for dislocation–solute interactions and to assess the consistency of the QM/ML coupling, MLIP-only energies were evaluated for different region selections both before and after the QM/ML production runs. In these MLIP-only calculations, the reference energies in equation

(3) for perfect hcp Zr with and without solutes were replaced by the corresponding MLIP values, ensuring internal consistency within the MLIP framework.

Given the multiregional structure of the model, a key validation is whether the dominant contribution to the solute–dislocation binding energies is effectively localised within the DFT zones (Zone I + II). To test this, binding energies were evaluated using three region definitions: the DFT zones (Zone I + II), an extended region including the ML embedding (Zone I + II + III), and the full model including the fixed outer buffer (Zone I + II + III + IV). Comparing MLIP-only binding energies across these region selections provides a practical diagnostic of the sensitivity of the interaction to model size.

If the binding energies remain consistent as the region size increases, this suggests that the dominant contribution is adequately captured within the DFT zones, and the QM/ML partition is expected to provide reliable energetics. In contrast, significant variation with increasing region size indicates either that the MLIP introduces artificial long-range solute–strain interactions, or that the physical interaction extends beyond the DFT zones and is therefore not fully captured by DFT. In both cases, such behaviour signals potential limitations in the accuracy of the QM/ML description.

To ensure that differences in binding energy arise solely from region selection rather than boundary inconsistencies, a consistent atom selection procedure was employed. For each dislocation configuration, the solute-free system was first relaxed, after which the solute was introduced into the relaxed structure. The atom indices defining the DFT zones and outer buffer were then fixed and reused for all corresponding solute-containing configurations. As a result, the geometry of the outer buffer remains unchanged between solute-free and solute-containing systems, allowing boundary contributions to cancel when evaluating binding energies.

As shown in Fig. 8, MLIP binding energies exhibit only modest variation between the different region selections for most configurations evaluated on the post-QM/ML relaxed structures. In particular, the binding energies obtained for the Zone I+II region remain broadly consistent with those evaluated for the larger Zone I+II+III and Zone I+II+III+IV regions, although the smallest Zone I+II region is expected to be most sensitive to finite-size effects and interactions with the artificial vacuum. In most cases, these differences remain limited, indicating that the MLIP contribution outside the DFT zones is reasonably well behaved and that the present QM/ML partition is generally appropriate.

A second comparison can be made between the pre-QM/ML pre-relaxed full-model (all four zones) and the post-QM/ML full-model MLIP binding energies. For H and Sn, these all-zone MLIP binding energies do not differ strongly, suggesting that the MLIP-based sampling procedure provides a reasonably reliable identification of representative solute sites. By contrast, larger deviations are observed for some of the high-energy Fe and Nb configurations, corresponding to the upper ends of the binding-energy distributions. In these cases, the MLIP binding energy for all regions can shift substantially after the QM/ML relaxation, in some instances even changing from attractive to repulsive. This indicates less reliable MLIP-only site stability predictions and suggests that purely MLIP-based sampling may underestimate the degree of repulsion in from dislocation cores. Overall, the presence of these more pronounced deviations shows that the observed model-size stability is not universal and must be assessed on a configuration-by-configuration basis.

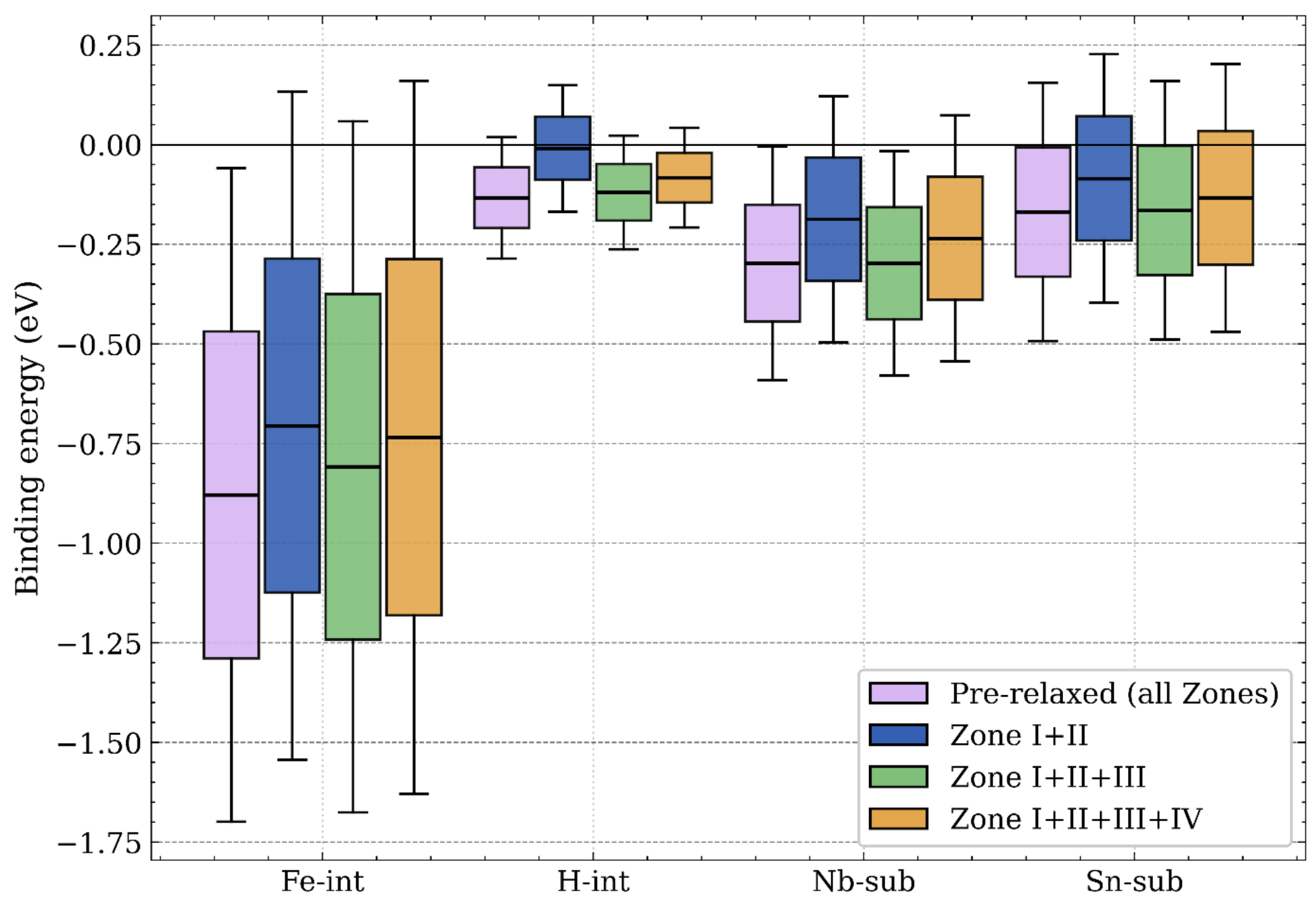

*Figure 8: MLIP-based binding energies for solutes binding to the <c/2> vacancy-type edge dislocation, evaluated from the pre-relaxed structures and from the MLIP component of the post-QM/ML structures. Post-QM/ML binding energies are evaluated and shown for different included regions: Zone I, II, III and IV. For a given solute–dislocation configuration, these energies are evaluated within the same QM/ML run.*

However, for certain configurations, deviations are more pronounced. For example, Sn binding to the edge segment of an ⟨a⟩ loop dislocation (Fig. 9) shows a significant difference between the binding energy evaluated within the DFT zones and that obtained when larger regions are included. This behaviour indicates increased sensitivity to model size and suggests either that the interaction is not sufficiently confined within the present DFT zones, or that MLIP-predicted long-range strain contributions make a non-negligible contribution to the energetics.

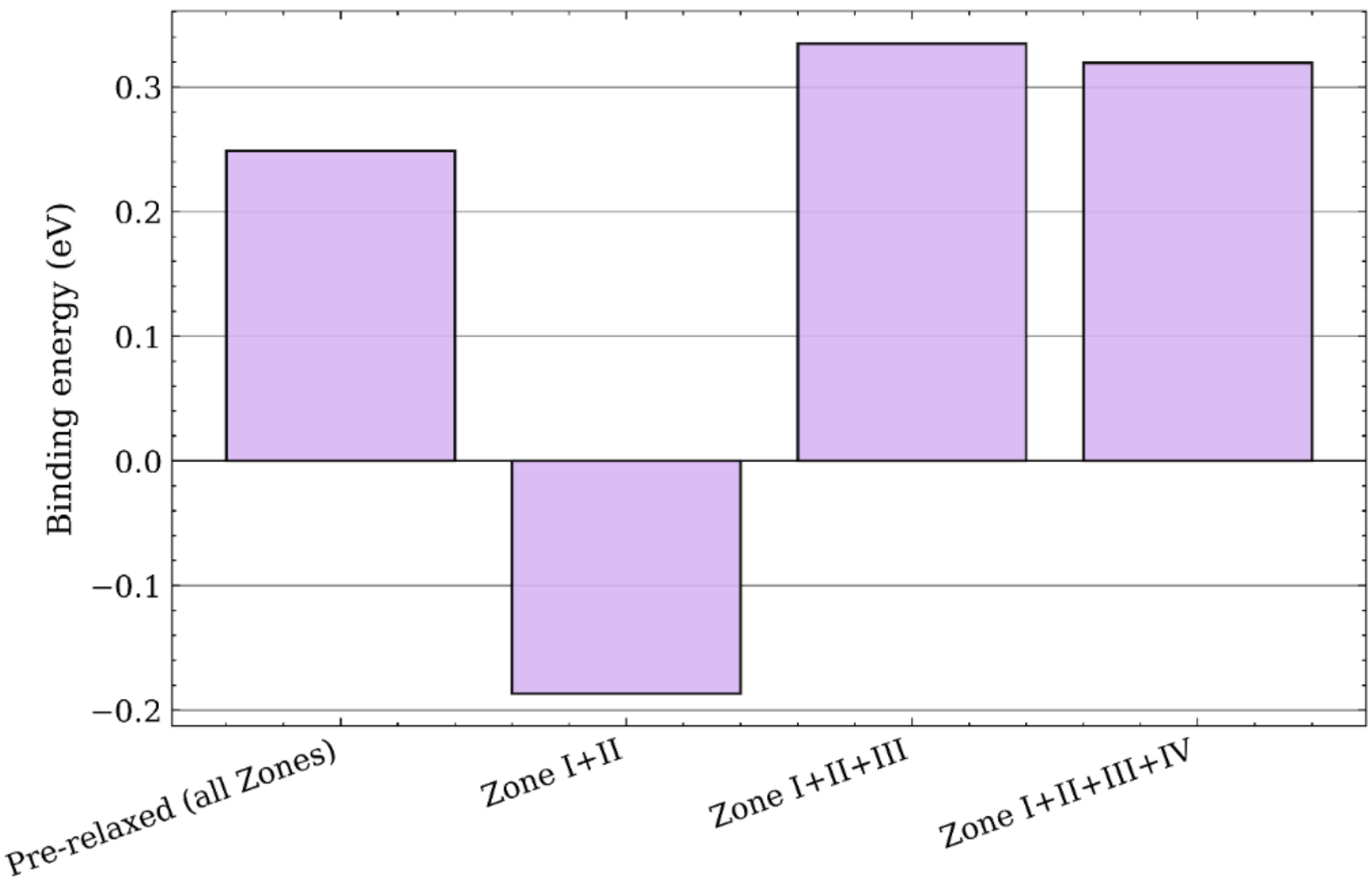

*Figure 9: MLIP-based binding energies for a Sn atom binding to the <a> loop edge segment, evaluated with different included regions.*

## Discussion

### Validity of the QM/ML correction

Sample calculations in both Zr and Fe show a consistent pattern: pure MLIP predictions tend to perform poorly for chemically sensitive dislocation–solute interactions, particularly for interstitial solutes, which are of direct relevance to irradiation-damage studies. The QM/ML framework was developed to address this limitation by restoring a DFT description in the defect-core region, while retaining a larger atomistic environment through the MLIP. In this way, the method aims to combine the short-range chemical accuracy of DFT with the long-range elastic embedding required for physically meaningful dislocation simulations.

The reliability of the framework depends on the distinct roles played by the DFT and MLIP components. Within the present implementation, the DFT region is intended to capture the dominant chemically sensitive contribution to the dislocation–solute interaction, including local bonding, charge redistribution and any defect-core electronic-structure effects that cannot be described reliably by the MLIP. By contrast, the surrounding MLIP regions provide the longer-range elastic response of the dislocation, stabilise the embedded relaxation of the DFT zones, and prevent unphysical boundary behaviour associated with the finite vacuum-containing geometry used in the DFT calculations. The method therefore does not require the total interaction to be fully local, but it does require the dominant DFT correction to be sufficiently local for the hybrid decomposition to remain meaningful.

This locality assumption can be examined directly through the region-dependent binding-energy analysis introduced in the present work. In the ideal case, binding energies evaluated using MLIP over Zone I+II, Zone I+II+III and Zone I+II+III+IV should remain broadly consistent, indicating that the dominant energetic contribution is already contained within the DFT zones and that the surrounding MLIP regions primarily provide mechanical embedding. However, when strong disagreement is observed between these region selections, the MLIP is effectively predicting a non-negligible energetic contribution outside the DFT zones, corresponding to a coupling between the local solute interaction and the surrounding dislocation strain field. In such cases, the QM/ML binding energy is no longer determined solely by the DFT-corrected core contribution, but also by the MLIP description of the outer region. While the formal energy expression remains valid, its physical interpretation as a local correction becomes less clear, and the result becomes sensitive to the accuracy of the MLIP outside the DFT region. It is generally not possible to determine whether this non-local contribution is physically meaningful or an artefact of the potential without substantially larger DFT calculations to compare to, which are impractical for systems of the present size. Two regimes can therefore be identified. In the favourable regime, the MLIP remains locally consistent with DFT within Zone I+II, and any non-local contribution is either small or reasonably described, so that the QM/ML correction remains meaningful. In the more problematic regime, the MLIP shows both strong region dependence and disagreement with DFT within Zone I+II. In this case, replacing the inner-region energy introduces an inconsistency between the DFT and MLIP contributions, and the resulting binding energy may become unreliable.

Overall, the QM/ML framework is most reliable when the DFT correction is effectively localised and the remaining MLIP contribution is either small or well-behaved. Configurations exhibiting strong region dependence should therefore be treated with caution, as they fall outside the regime in which the present approach can be expected to yield robust quantitative binding energies.

### Recovery of missing local physics and broader applicability

In addition to the locality requirement, the overall accuracy of the QM/ML framework depends on the quality of the underlying MLIP. Key factors include accurate defect formation energies, reliable elastic properties, and accurate lattice parameters. While the first two can often be partially corrected by the DFT contribution, the latter present more fundamental challenges. In particular, a lattice-parameter mismatch exceeding approximately 1% introduces an artificial strain into the DFT region, which can bias the calculated binding energies. This effect is especially pronounced for interstitial solutes, where local strain fields can partially compensate or amplify the imposed mismatch. Therefore, to use this method well, it requires good elastic constants and lattice parameter match of the MLIP to the DFT.

Beyond energetic accuracy, a central advantage of including a DFT region is the recovery of electronic-structure information absent from MLIP simulations. In hcp Zr, previous work has shown that Fe interstitials lose their magnetic moment due to strong hybridisation with neighbouring Zr atoms, whereas substitutional Fe remains magnetised because of weaker hybridisation and longer bond distances [31]. In dislocation cores, however, local geometries deviate significantly from bulk configurations, and partial recovery of magnetic moments cannot be excluded. Under these conditions, the validity of non-magnetic MLIPs becomes uncertain. In the present QM/ML framework, this issue can be assessed using projected density of states (PDOS) analysis. In non-spin-polarised calculations, the appearance of a sole broad peak centred at the Fermi level in the d orbital Fe-projected DOS indicates a magnetic instability and suggests that spin polarisation should be included [72]. This provides a practical diagnostic for identifying configurations in which additional electronic degrees of freedom must be considered.

The challenge is even greater in BCC Fe, where magnetism is intrinsic to the host lattice, and must be treated with spin polarisation to reproduce the correct chemistry and defect energetics. Fully DFT-based dislocation–solute calculations in Fe are therefore extremely expensive and difficult to reproduce [4,32,34,37], particularly when system sizes exceed one thousand atoms. Here, the QM/ML framework offers a practical route forward by reducing the number of DFT atoms from more than 1400 to approximately 300, while still embedding them within a realistic long-range elastic field supplied by the MLIP. This substantially lowers the computational cost and makes systematic spin-polarised dislocation–solute calculations feasible within realistic resources, enabling DFT-quality benchmark data to be generated for defect configurations that would otherwise remain inaccessible.

Taken together, these aspects define the main methodological advance of the present work: a transferable QM/ML framework that combines modern neural-network MLIPs with a multiregional energy decomposition and an explicit locality diagnostic. By reinstating a DFT description at the defect core, the method restores local chemistry and electronic structure, while the surrounding MLIP regions provide the long-range elastic embedding. At the same time, the region-dependent energy analysis provides a practical criterion for assessing the validity of the hybrid partition. This allows the framework not only to generate high-fidelity reference data, but also to identify configurations in which the present approach is less reliable and where more advanced modelling is required. More broadly, as a non-empirical first-principles framework, the method provides a predictive route for exploring new alloy systems and for computational characterisation prior to experimental investigation.

## Physical interpretation of the sample results

The solute–dislocation binding energies obtained from the QM/ML framework reproduce trends that are consistent with existing atom probe tomography (APT) and transmission electron microscopy (TEM) observations [27,30]. Fe solutes display a strong preference to segregate to dislocation cores in Zr alloys, whereas Sn shows a weakly repulsive interaction. Combined with the high diffusivity of Fe and the previously reported repulsion between Sn and Fe in atomistic simulation [73], this supports the view that slowly diffusing Sn is unlikely to accumulate at dislocation loops, while the more mobile Fe readily migrates to and binds strongly with them. Pure universal MLIP predictions tend, however, to underestimate the degree of Sn repulsion from the <a> loop screw core, reinforcing the need for a QM-corrected description.

Nb exhibits a slight attractive interaction with dislocation cores in the present QM/ML calculations. However, its relatively low diffusivity suggests that direct segregation to dislocation loops is kinetically limited. This behaviour appears inconsistent with experimental observations, which show that Nb is not strongly associated with loop decoration but instead is enriched in planar features between loops [30]. Previous atomistic studies of solute clustering in perfect crystals have shown that, while Fe and Sn exhibit a mutual repulsion, Nb does not strongly repel either species and can reduce the effective binding energy of Fe/Sn clusters towards zero [73]. Based on these observations, we propose that the role of Nb in irradiated Zr alloys is not primarily to bind directly to dislocation cores, but rather to stabilise Sn-rich planar configurations that form in the surrounding microstructure. In this interpretation, the apparent anticorrelation between Fe-decorated dislocation loops and Nb/Sn-rich planar regions arises from a combination of thermodynamic and kinetic effects: the high diffusivity of Fe promotes rapid segregation to dislocations, while the slower diffusion of Nb favours its incorporation into more stable, co-segregated structures with Sn away from the dislocation core.

Beyond interpreting these specific trends, the results demonstrate that the QM/ML framework provides a practical route for benchmarking fine-tuned NN-based MLIPs. While the present study focuses on single-solute binding to <a> and <c/2> loop segments, the same approach can be extended, in line with previous QM/MM work, to more complex defect phenomena [53-55, 58]. These include solute clustering around dislocation cores, solute–solute interactions mediated by dislocation strain fields, solute diffusion across dislocations and the influence of solutes on the dislocation Peierls barrier. BCC Fe results from the QM/ML code are also in line with the previous pure DFT simulations results that show similar binding of Sn and C to the dislocations [4, 34].

The QM/ML code can be used for better understanding the microstructural evolution and mechanical properties of the structural materials not limited by the nuclear application. For instance, it can be applied to quantify solute-induced dislocation pinning, changes in dislocation mobility and stability, and the resulting impact on strain localization and crack initiation [53-55,74]. Consequently, the framework not only reproduces experimentally relevant segregation behaviour but also offers a scalable computational tool for studying and validating solute–dislocation interaction mechanisms across a broad range of irradiation-damaged alloy systems.

# Conclusions

1. We have developed a neural network MLIP based force convergence hybrid QM/ML framework specifically for dislocation studies. By combining the accuracy of DFT with the efficiency of modern MLIPs, the method reduces the required DFT dislocations calculations from many thousands of atoms to only a few hundred, thereby significantly lowering computational cost.
2. We demonstrate the necessary benchmarking procedures for the multiregion disk model and show that the QM/ML approach corrects MLIP only predictions of solute–dislocation binding energies. The inclusion of a DFT core also recovers the electronic-structure information that is absent in pure MLIP simulations, providing a more fundamental atomistic understanding of the interactions.
3. The QM/ML framework can be used to quantify and visualise sources of error in MLIP defect simulations and offers practical guidance for benchmarking and improving fine-tuned potentials. When combined with standard validation tests, this makes the origin of MLIP inaccuracies and the areas requiring further refinement much clearer.
4. For Zr systems, the sample results reproduce experimentally observed trends, including strong segregation of Fe to dislocation cores and the anticorrelated Sn segregation relative to Fe. This demonstrates the capability of the framework to capture realistic solute–dislocation interaction mechanisms in irradiated alloys.

# Method

## Computational details

The QM/ML simulations were performed using a combined framework in which the quantum-mechanical region was evaluated with the Vienna Ab initio Simulation Package (VASP) [60,61]. The projector-augmented wave (PAW) method with the Perdew–Burke–Ernzerhof (PBE) exchange–correlation functional was employed throughout [62,63]. A plane-wave cutoff of 400 eV and Methfessel–Paxton smearing with a width of 0.2 eV were applied [64]. Electronic convergence thresholds were set to $1\times10^{-6}$ eV for Zr host calculations and $1\times10^{-5}$ eV for BCC Fe calculations. Real-space projection of the PAW projectors was used for all calculations. Spin polarisation was included for all BCC Fe systems with an initial magnetic moment of 2.8 $\mu B$ on all Fe atoms.

Periodic boundary conditions were applied only along the dislocation-line direction in the MLIP calculations, with vacuum regions introduced in the nonperiodic directions. Accordingly, a single k-point was used for the vacuum containing directions in each DFT calculation, and a mesh of 6 k points was therefore selected for the periodic direction. The PAW datasets employed were: Zr ($4s^2$ $4p^6$ $4d^3$ $5s^1$), Nb ($4p^6$ $4d^4$ $5s^1$), Sn ($5s^2$ $5p^2$), Fe ($3d^7$ $4s^1$) and H ($1s^1$), where the atomic orbitals in the brackets are those treated as valance states.

All VASP and MLIPs calculations were executed with the Atomistic Simulation Environment (ASE) [65]. For MLIP calculations, structural relaxations were performed using the fast inertial relaxation engine (FIRE) algorithm [66]. A force convergence criterion of 0.01 eV $Å^{-1}$ (maximum force component) was used for MLIP-only pre-relaxation, and 0.05 eV $Å^{-1}$ for the subsequent QM/ML production runs. A restart handler was incorporated to read ASE trajectory files and continue simulations seamlessly from the final step of previous runs. Periodic boundary conditions for MLIP calculations were defined consistently with the VASP setup.

For Zr-based systems, the medium-sized MACE-MPA-0 model (float-32 precision) was used to balance accuracy and computational efficiency. For BCC Fe, a sample NEP-ft was employed [67]. This NEP was fine-tuned from NEP89 [68] using the GPUMD software [69] on a dataset comprising 18 lattice-deformation paths and 21 distinct point-defect configurations in pure BCC Fe. This NEP-ft potential was designed to correct the NEP89 errors in the elastic contribution, but not modify the chemical contribution, to the solute binding energies.

## Overview of QM/ML framework

A central challenge in QM/MM-type simulations is the consistent treatment of forces and energies across the interface between different levels of theory [53–55]. In the present QM/ML framework, this issue is addressed using a four-region cylindrical (disk-shaped) model, illustrated in Fig. 10, which enables seamless coupling between DFT and modern MLIPs.

From the centre outward, Zone I defines the QM region, which contains both the solute and the dislocation line, the latter aligned with the cylinder axis. Energies and forces in this region are evaluated fully at the DFT level to capture the solute chemistry and electronically complex defect core. Surrounding Zone I, Zone II acts as an inner buffer (IB) region. This region is treated using a hybrid description, combining DFT energies with MLIP forces, and serves to shield the QM region from the artificial vacuum introduced in the DFT calculations. The thickness of Zone II is determined through convergence testing to ensure accurate force transfer and to minimise free-boundary artefacts.

Zones III and IV, referred to as the ML region and outer buffer region respectively, are described entirely using MLIP energies and forces. These outer regions provide the extended atomistic environment required to reproduce the long-range elastic field of the dislocation and any associated stacking-fault geometries. They also ensure physically meaningful relaxation of the inner regions by preventing unphysical distortions of the DFT zones (Zone I+II) that would otherwise arise from the presence of vacuum.

A vacuum layer of 30 Å is introduced beyond Zone IV. Due to the proximity of outer atoms to vacuum, boundary atoms exhibit artificially elevated energies and forces. To prevent these edge effects from propagating into the central regions, atoms within Zone IV are held fixed during the QM/ML relaxation following an initial full-MLIP pre-relaxation. The dislocation configurations were generated using Atomsk, and structural analysis and visualisation were performed using OVITO [70,71].

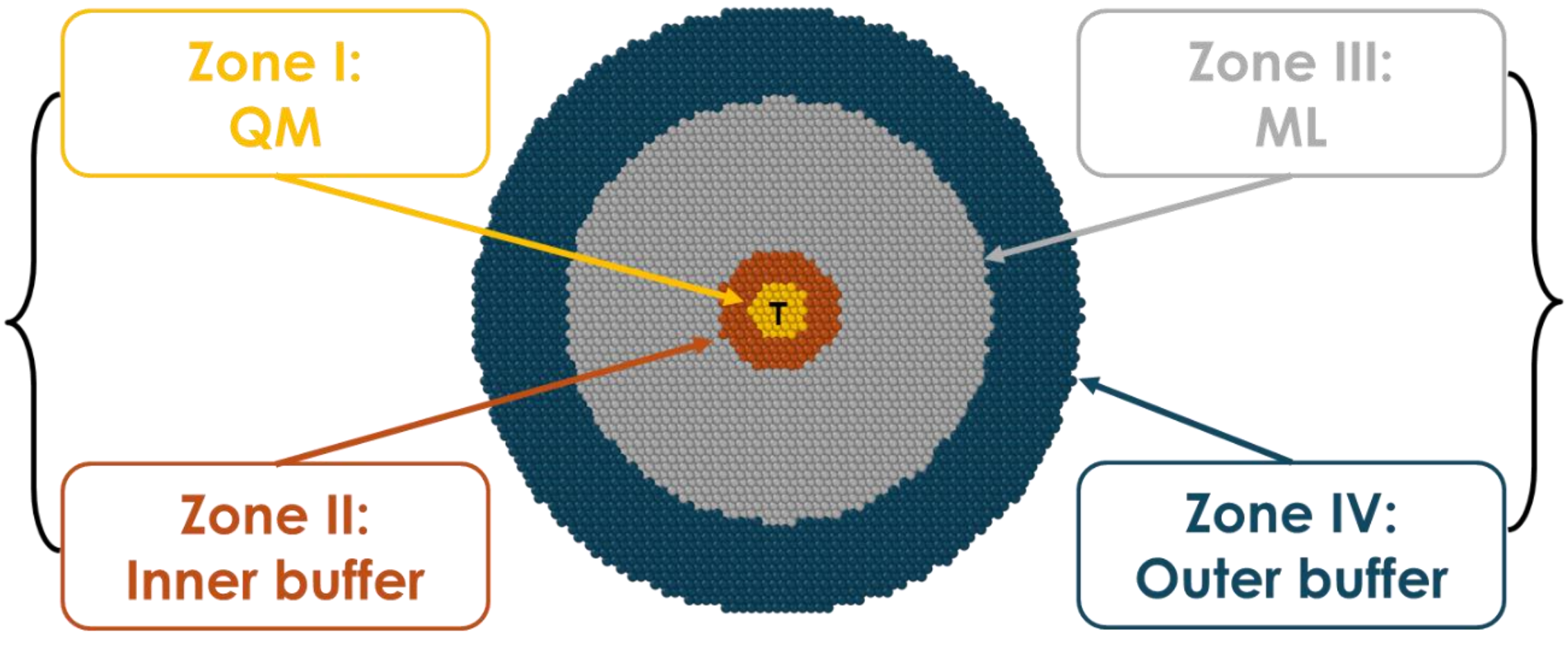

*Figure 10: Multiregional model structure used in the QM/ML framework.*

The QM/ML workflow was implemented within the Python-based Atomistic Simulation Environment (ASE), using DFT (VASP package) and the MLIPs as independent ASE calculators (Fig. 11). The procedure begins with an MLIP-only structural pre-relaxation FIRE algorithm [66], following which the four model regions are defined. After pre-relaxation, separate DFT and MLIP calculations are performed to obtain energies and forces for the relevant regions. These are then combined into system-level QM/ML energies and forces using the following expressions:

$$E_{QM/ML}^{I+II+III+IV} = E_{MLIP}^{I+II+III+IV} + E_{DFT}^{I+II} - E_{MLIP}^{I+II} \quad (1)$$

$$\vec{F}_{QM/ML}^{I+II+III+IV} = \vec{F}_{MLIP}^{I+II+III+IV} + \vec{F}_{DFT}^{I} - \vec{F}_{MLIP}^{I} \quad (2)$$

The resulting mixed energies and forces are transferred back to ASE for relaxation with FIRE until force convergence is reached. In addition to the QM/ML energies, the code outputs the individual MLIP and DFT contributions for different regions, enabling detailed comparison between potentials and benchmarking of the hybrid scheme.

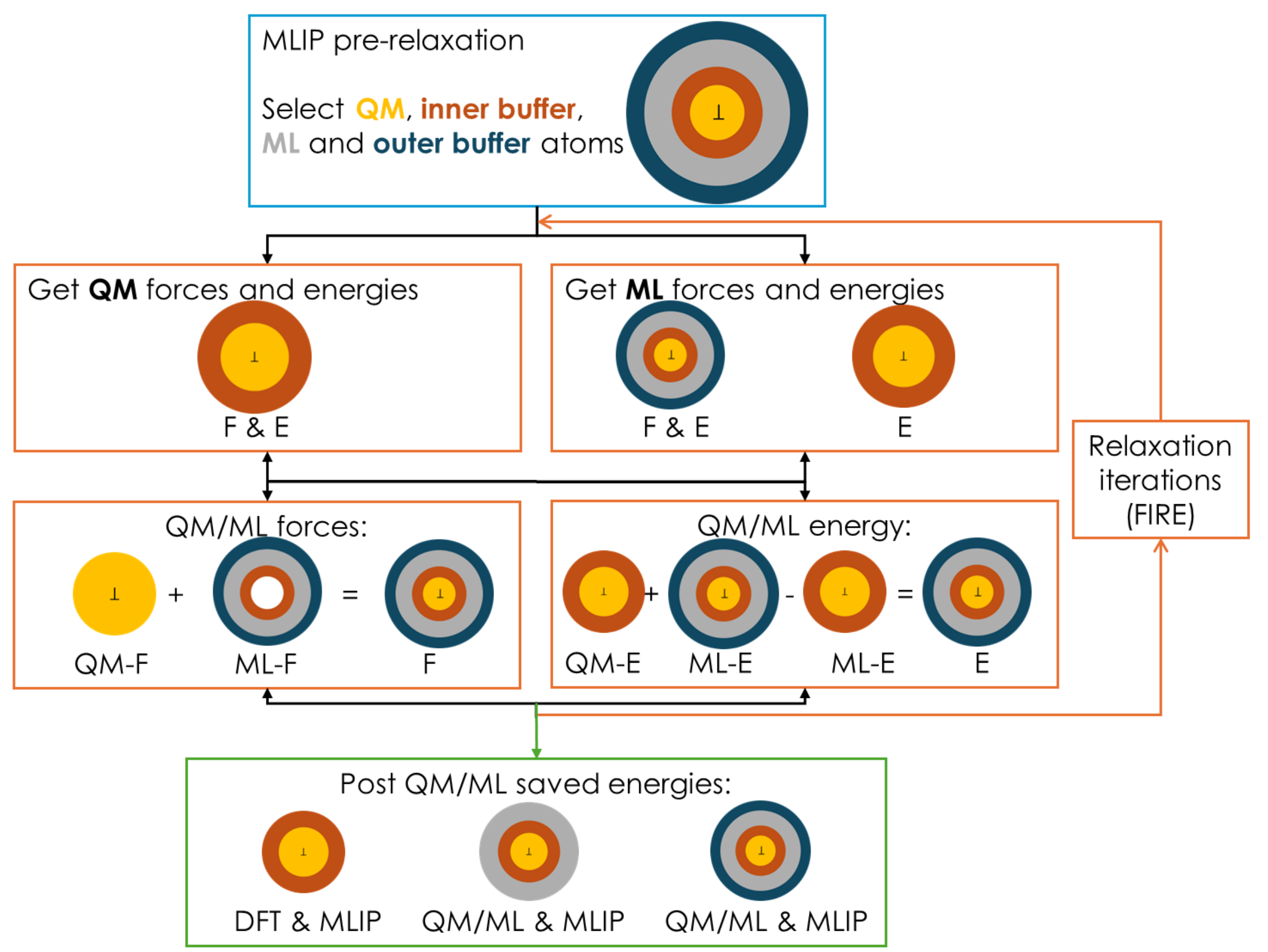

*Figure 11: QM/ML calculation workflow. The model regions are defined the same as in Fig 10.*

### Binding energy evaluation

Taking Zr systems as an example, the solute–dislocation binding energy in the QM/ML framework is evaluated as,

$$E_{QM/ML}^{binding} = (E_{QM/ML}^{dislo+sol} - E_{QM/ML}^{dislo}) - (E_{DFT}^{Zr+sol} - E_{DFT}^{Zr}) \quad (3)$$

Thus, two QM/ML simulations are required: a dislocation containing the solute ($E_{QM/ML}^{dislo+sol}$) and the corresponding solute-free reference ($E_{QM/ML}^{dislo}$). The DFT energy difference for a solute in a perfect hcp Zr cell is subtracted to isolate the interaction energy between the dislocation core and the solute. This corresponds to a solute moving from somewhere in the Zr matrix, but far from the dislocation to a position near the dislocation.

Binding energies are computed using different combinations of regions and mixing criteria (See Supplementary Information). Including variants that use either DFT or MLIP reference energies for the corresponding perfect structures, enabling systematic benchmarking of MLIP performance.

## Acknowledgement

This work was performed using resources provided by the Cambridge Service for Data Driven Discovery (CSD3) operated by the University of Cambridge Research Computing Service (www.csd3.cam.ac.uk), provided by Dell EMC and Intel using Tier-2 funding from the Engineering and Physical Sciences Research Council (capital grant EP/T022159/1), and DiRAC funding from the Science and Technology Facilities Council (www.dirac.ac.uk). We are grateful to the UK Materials and Molecular Modelling Hub for computational resources, which is partially funded by EPSRC (EP/T022213/1, EP/W032260/1 and EP/P020194/1). We acknowledge computational resources and support provided by the Imperial College Research Computing Service (http://doi.org/10.14469/hpc/2232). The authors would like to thank the Engineering and Physical Sciences Research Council UK for funding the study through the MIDAS (Mechanistic understanding of Irradiation Damage in fuel Assemblies) programme grant (EP/S01702X/1). ChatGPT was used to assist with grammatical polishing, improving readability and providing guidance during the development of the Python code.

## Data availability

All data generated or analysed during this study are available from the corresponding author on reasonable request.

## Code availability

The code supporting the findings of this study will be made publicly available on GitHub upon publication at https://github.com/antony0429/Hybrid-QM-ML-framework .

## References:

1. J. P. Hirth and J. Lothe, Theory of Dislocations, 2nd ed. Wiley, 1982.
2. J. Koutský and J. Kocík, Radiation Damage of Structural Materials. Amsterdam: Elsevier Science, 2014.
3. M. Griffiths, "A review of microstructure evolution in zirconium alloys during irradiation," Journal of Nuclear Materials, vol. 159, pp. 190–218, Oct. 1988. doi:10.1016/0022-3115(88)90093-1
4. C. Domain and C. S. Becquart, "Solute – 〈111〉 interstitial loop interaction in α-FE: A DFT study," Journal of Nuclear Materials, vol. 499, pp. 582–594, Feb. 2018. doi:10.1016/j.jnucmat.2017.10.070
5. D. Terentyev, F. Bergner, and Y. Osetsky, "Cr segregation on dislocation loops enhances hardening in ferritic Fe–Cr Alloys," Acta Materialia, vol. 61, no. 5, pp. 1444–1453, Mar. 2013. doi:10.1016/j.actamat.2012.11.021
6. S. C. Lumley et al., "The stability of alloying additions in zirconium," Journal of Nuclear Materials, vol. 437, no. 1–3, pp. 122–129, Jun. 2013. doi:10.1016/j.jnucmat.2013.01.335
7. P. A. Burr, S. T. Murphy, S. C. Lumley, M. R. Wenman, and R. W. Grimes, "Hydrogen solubility in zirconium intermetallic second phase particles," Journal of Nuclear Materials, vol. 443, no. 1–3, pp. 502–506, Nov. 2013. doi:10.1016/j.jnucmat.2013.07.060
8. W. J. S. Yang, R. P. Tucker, B. Cheng, and R. B. Adamson, "Precipitates in zircaloy: Identification and the effects of irradiation and thermal treatment," Journal of Nuclear Materials, vol. 138, no. 2–3, pp. 185–195, Apr. 1986. doi:10.1016/0022-3115(86)90005-x
9. M. Griffiths, R. W. Gilbert, and G. J. C. Carpenter, "Phase instability, decomposition and redistribution of intermetallic precipitates in zircaloy-2 and -4 during neutron irradiation," Journal of Nuclear Materials, vol. 150, no. 1, pp. 53–66, Sep. 1987. doi:10.1016/0022-3115(87)90093-6
10. D. O. Pickman, "Interactions between fuel pins and assembly components," Nuclear Engineering and Design, vol. 33, no. 2, pp. 125–140, Sep. 1975. doi:10.1016/0029-5493(75)90018-7
11. M. Griffiths, R. A. Holt, and A. Rogerson, "Microstructural aspects of accelerated deformation of Zircaloy nuclear reactor components during service," Journal of Nuclear Materials, vol. 225, pp. 245–258, Aug. 1995. doi:10.1016/0022-3115(94)00687-3
12. R. A. Holt and R. W. Gilbert, "〈c〉 component dislocations in annealed zircaloy irradiated at about 570 K," Journal of Nuclear Materials, vol. 137, no. 3, pp. 185–189, Feb. 1986. doi:10.1016/0022-3115(86)90218-7

13. V. Shishov et al., “Influence of neutron irradiation on dislocation structure and phase composition of Zr-Base Alloys,” Zirconium in the Nuclear Industry: Eleventh International Symposium, pp. 603–622, Jan. 1996. doi:10.1520/stp16192s
14. J. Ribis et al., “Experimental and modeling approach of irradiation defects recovery in zirconium alloys: Impact of an applied stress,” Zirconium in the Nuclear Industry: 15th International Symposium, pp. 674–695, Jan. 2009. doi:10.1520/stp48162s
15. E. Wakai et al., “Irradiation damages of structural materials under different irradiation environments,” Journal of Nuclear Materials, vol. 543, p. 152503, Jan. 2021. doi:10.1016/j.jnucmat.2020.152503
16. M. Nastar and F. Soisson, “Radiation-induced segregation,” Comprehensive Nuclear Materials, pp. 471–496, 2012. doi:10.1016/b978-0-08-056033-5.00035-5
17. J. A. Horak and H. V. Rhude, “Irradiation growth of zirconium — plutonium alloys,” Journal of Nuclear Materials, vol. 3, no. 1, pp. 111–112, Jan. 1961. doi:10.1016/0022-3115(61)90185-4
18. A. C. Jain, P. A. Burr, and D. R. Trinkle, “First-principles calculations of solute transport in zirconium: Vacancy-mediated diffusion with metastable states and interstitial diffusion,” Physical Review Materials, vol. 3, no. 3, Mar. 2019. doi:10.1103/physrevmaterials.3.033402
19. F. Garzarolli, H. Stehle, and E. Steinberg, “Behavior and properties of Zircaloys in power reactors: A short review of pertinent aspects in LWR Fuel,” Zirconium in the Nuclear Industry: Eleventh International Symposium, pp. 12–32, Jan. 1996. doi:10.1520/stp16165s
20. H. G. Rickover, L. D. Geiger, and B. Lustman, History of the development of zirconium alloys for use in nuclear reactors, Mar. 1975. doi:10.2172/4240391
21. T. Fujita, J. Hirabayashi, Y. Katayama, F. Kano, and H. Watanabe, “Contribution of dislocation loop to radiation-hardening of RPV steels studied by STEM/EDS with surveillance test pieces,” Journal of Nuclear Materials, vol. 572, p. 154055, Dec. 2022. doi:10.1016/j.jnucmat.2022.154055
22. K. Inoue et al., “Effects of P on formation and growth of Mn-Ni-Si clusters in low-cu reactor pressure vessel steel analyzed by Atom Probe Tomography,” Acta Materialia, vol. 296, p. 121303, Sep. 2025. doi:10.1016/j.actamat.2025.121303
23. Q. Yuan et al., In-situ S/TEM investigations of deformation and damage mechanism of neutron-irradiated eurofer97, 2024. doi:10.2139/ssrn.4812461
24. G. J. C. Carpenter, R. H. Zee, and A. Rogerson, “Irradiation growth of zirconium single crystals: A Review,” Journal of Nuclear Materials, vol. 159, pp. 86–100, Oct. 1988. doi:10.1016/0022-3115(88)90087-6
25. A. Harte et al., “The effect of matrix chemistry on dislocation evolution in an irradiated Zr Alloy,” Acta Materialia, vol. 130, pp. 69–82, May 2017. doi:10.1016/j.actamat.2017.03.024
26. D. Mayweg et al., Formation of pure zirconium islands inside C-component loops in high-burnup fuel cladding, 2024. doi:10.2139/ssrn.4754309
27. G. Sundell et al., “Redistribution of alloying elements in zircaloy-2 after in-reactor exposure,” Journal of Nuclear Materials, vol. 454, no. 1–3, pp. 178–185, Nov. 2014. doi:10.1016/j.jnucmat.2014.07.072
28. T. Sawabe and T. Sonoda, “Evolution of nanoscopic iron clusters in irradiated zirconium alloys with different iron contents,” Journal of Nuclear Science and Technology, vol. 55, no. 10, pp. 1110–1118, May 2018. doi:10.1080/00223131.2018.1479987
29. J. Eriksson, G. Sundell, P. Tejland, H.-O. Andrén, and M. Thuvander, “An atom probe tomography study of the chemistry of radiation-induced dislocation loops in zircaloy-2 exposed to boiling water reactor operation,” Journal of Nuclear Materials, vol. 550, p. 152923, Jul. 2021. doi:10.1016/j.jnucmat.2021.152923
30. B. M. Jenkins, J. Haley, M. P. Moody, J. M. Hyde, and C. R. M. Grovenor, “APT and TEM study of behaviour of alloying elements in neutron-irradiated zirconium-based alloys,” Scripta Materialia, vol. 208, p. 114323, Feb. 2022. doi:10.1016/j.scriptamat.2021.114323

31. J. Zhang, A. Horsfield, and M. Wenman, “Density functional theory simulation study of FE solutes in HCP zirconium: Magnetic and electronic properties,” Journal of Nuclear Materials, vol. 609, p. 155755, May 2025. doi:10.1016/j.jnucmat.2025.155755
32. C. Woodward, D. R. Trinkle, L. G. Hector, and D. L. Olmsted, “Prediction of dislocation cores in aluminum from density functional theory,” Physical Review Letters, vol. 100, no. 4, Jan. 2008. doi:10.1103/physrevlett.100.045507
33. R. Hulse and C. P. Race, “An atomistic modelling study of the properties of dislocation loops in zirconium,” Journal of Nuclear Materials, vol. 546, p. 152752, Apr. 2021. doi:10.1016/j.jnucmat.2020.152752
34. R. Candela, N. Mousseau, R. G. Veiga, C. Domain, and C. S. Becquart, “Interaction between interstitial carbon atoms and a 1/2 ⟨111⟩ self-interstitial atoms loop in an iron matrix: A combined DFT, off lattice KMC and MD study,” Journal of Physics: Condensed Matter, vol. 30, no. 33, p. 335901, Jul. 2018. doi:10.1088/1361-648x/aad25d
35. C. Varvenne, O. Mackain, L. Proville, and E. Clouet, “Hydrogen and vacancy clustering in zirconium,” Acta Materialia, vol. 102, pp. 56–69, Jan. 2016. doi:10.1016/j.actamat.2015.09.019
36. R. O. Jones, “Density functional theory: Its origins, rise to prominence, and future,” Reviews of Modern Physics, vol. 87, no. 3, pp. 897–923, Aug. 2015. doi:10.1103/revmodphys.87.897
37. J. F. March-Rico, G. Huang, and B. D. Wirth, “The effect of local chemical environment on the Energetics of stacking faults and vacancy platelets in α-Zirconium,” Journal of Nuclear Materials, vol. 540, p. 152339, Nov. 2020. doi:10.1016/j.jnucmat.2020.152339
38. J. F. March-Rico and B. D. Wirth, “Stress states and point defect capture radii of dislocation a-loops and C-loops in alpha-zirconium,” Journal of Nuclear Materials, vol. 587, p. 154752, Dec. 2023. doi:10.1016/j.jnucmat.2023.154752
39. D. C. Parfitt, C. L. Bishop, M. R. Wenman, and R. W. Grimes, “Strain fields and line energies of dislocations in uranium dioxide,” Journal of Physics: Condensed Matter, vol. 22, no. 17, p. 175004, Apr. 2010. doi:10.1088/0953-8984/22/17/175004
40. A. A. Kohnert, M. A. Cusentino, and B. D. Wirth, “Molecular statics calculations of the biases and point defect capture volumes of small cavities,” Journal of Nuclear Materials, vol. 499, pp. 480–489, Feb. 2018. doi:10.1016/j.jnucmat.2017.12.005
41. C. Dai et al., “Stability of vacancy and interstitial dislocation loops in α-zirconium: Atomistic calculations and continuum modelling,” Journal of Nuclear Materials, vol. 554, p. 153059, Oct. 2021. doi:10.1016/j.jnucmat.2021.153059
42. M. S. Daw and M. I. Baskes, “Embedded-atom method: Derivation and application to impurities, surfaces, and other defects in metals,” Physical Review B, vol. 29, no. 12, pp. 6443–6453, Jun. 1984. doi:10.1103/physrevb.29.6443
43. S. Zhang, Y. Li, S. Suzuki, A. Nakamura, and S. Ogata, “Neural network potential for dislocation plasticity in Ceramics,” npj Computational Materials, vol. 10, no. 1, Nov. 2024. doi:10.1038/s41524-024-01456-7
44. T. Okita et al., “Construction of machine-learning ZR interatomic potentials for identifying the formation process of C-type dislocation loops,” Computational Materials Science, vol. 202, p. 110865, Feb. 2022. doi:10.1016/j.commatsci.2021.110865
45. L. Mismetti and M. Hodapp, “Automated atomistic simulations of dissociated dislocations with ab initio accuracy,” Physical Review B, vol. 109, no. 9, Mar. 2024. doi:10.1103/physrevb.109.094120
46. J. Byggmästar, A. Hamedani, K. Nordlund, and F. Djurabekova, “Machine-learning interatomic potential for radiation damage and defects in Tungsten,” Physical Review B, vol. 100, no. 14, Oct. 2019. doi:10.1103/physrevb.100.144105
47. Z. Fan et al., General-purpose machine-learned potential for 16 elemental metals and their alloys, Dec. 2023. doi:10.21203/rs.3.rs-3612294/v1

48. L. Zhang, G. Csányi, E. van der Giessen, and F. Maresca, "Efficiency, accuracy, and transferability of machine learning potentials: Application to dislocations and cracks in iron," Acta Materialia, vol. 270, p. 119788, May 2024. doi:10.1016/j.actamat.2024.119788
49. R. Drautz, "Atomic cluster expansion for accurate and transferable interatomic potentials," Physical Review B, vol. 99, no. 1, Jan. 2019. doi:10.1103/physrevb.99.014104
50. I. Batatia, D. P. Kovács, G. N. C. Simm, C. Ortner, and G. Csányi, "MACE: Higher Order Equivariant Message Passing Neural Networks for Fast and Accurate Force Fields," arXiv preprint arXiv:2206.07697, Jun. 2022. doi:10.48550/arXiv.2206.07697.
51. G. Wang et al., "Machine learning interatomic potential: Bridge the gap between small-scale models and realistic device-scale simulations," iScience, vol. 27, no. 5, p. 109673, May 2024. doi:10.1016/j.isci.2024.109673
52. J. Kan et al., "A machine learning potential for simulation the dislocation behavior of magnesium," Journal of Magnesium and Alloys, Nov. 2024. doi:10.1016/j.jma.2024.11.009
53. Y. Zhao and G. Lu, "QM/MM study of dislocation—hydrogen/helium interactions in α-FE," Modelling and Simulation in Materials Science and Engineering, vol. 19, no. 6, p. 065004, Jul. 2011. doi:10.1088/0965-0393/19/6/065004
54. P. Grigorev, T. D. Swinburne, and J. R. Kermode, "Hybrid Quantum/classical study of hydrogen-decorated screw dislocations in tungsten: Ultrafast pipe diffusion, core reconstruction, and effects on glide mechanism," Physical Review Materials, vol. 4, no. 2, Feb. 2020. doi:10.1103/physrevmaterials.4.023601
55. T. D. Swinburne and J. R. Kermode, "Computing energy barriers for rare events from hybrid quantum/classical simulations through the virtual work principle," Physical Review B, vol. 96, no. 14, Oct. 2017. doi:10.1103/physrevb.96.144102
56. A. Marefat Khah, P. Reinholdt, J. M. Olsen, J. Kongsted, and C. Hättig, "Avoiding electron spill-out in QM/mm calculations on excited states with simple pseudopotentials," Journal of Chemical Theory and Computation, vol. 16, no. 3, pp. 1373–1381, Feb. 2020. doi:10.1021/acs.jctc.9b01162
57. J. Ho, H. Yu, Y. Shao, M. Taylor, and J. Chen, "How accurate are QM/MM models?," The Journal of Physical Chemistry A, vol. 129, no. 6, pp. 1517–1528, Dec. 2024. doi:10.1021/acs.jpca.4c06521
58. P. Grigorev, A. M. Goryaeva, M.-C. Marinica, J. R. Kermode, and T. D. Swinburne, "Calculation of dislocation binding to helium-vacancy defects in tungsten using hybrid ab initio-machine learning methods," Acta Materialia, vol. 247, p. 118734, Apr. 2023. doi:10.1016/j.actamat.2023.118734
59. Y.-J. Zhang, A. Khorshidi, G. Kastlunger, and A. A. Peterson, "The potential for machine learning in hybrid QM/mm calculations," The Journal of Chemical Physics, vol. 148, no. 24, Jun. 2018. doi:10.1063/1.5029879
60. J. Hafner, "ab-initio simulations of materials using VASP: Density-functional theory and beyond," Journal of Computational Chemistry, vol. 29, no. 13, pp. 2044–2078, 2008.
61. G. Kresse and J. Furthmüller, "Efficiency of AB-initio total energy calculations for metals and semiconductors using a plane-wave basis set," Computational Materials Science, vol. 6, no. 1, pp. 15–50, 1996.
62. G. Kresse and D. Joubert, "From Ultrasoft pseudopotentials to the projector augmented-wave method," Physical Review B, vol. 59, no. 3, pp. 1758–1775, 1999.
63. J. P. Perdew, K. Burke, and M. Ernzerhof, "Generalized gradient approximation made simple," Physical Review Letters, vol. 77, no. 18, pp. 3865–3868, 1996.
64. M. Methfessel, A.T. Paxton, High-precision sampling for Brillouin-zone integration in metals, Phys. Rev. B. 40 (1989) 3616–3621.
65. A. Hjorth Larsen et al., "The atomic simulation environment—a python library for working with atoms," Journal of Physics: Condensed Matter, vol. 29, no. 27, p. 273002, Jun. 2017. doi:10.1088/1361-648x/aa680e

66. E. Bitzek, P. Koskinen, F. Gähler, M. Moseler, and P. Gumbsch, "Structural relaxation made simple," Physical Review Letters, vol. 97, no. 17, Oct. 2006. doi:10.1103/physrevlett.97.170201
67. Z. Fan et al., "Neuroevolution machine learning potentials: Combining high accuracy and low cost in atomistic simulations and application to heat transport," Physical Review B, vol. 104, no. 10, Sep. 2021. doi:10.1103/physrevb.104.104309
68. J. Xu et al., NEP89: Universal neuroevolution potential for inorganic and organic materials across 89 elements, Aug. 2025. doi:10.21203/rs.3.rs-7058411/v1
69. Z. Fan et al., "GPUMD: A package for constructing accurate machine-learned potentials and performing highly efficient atomistic simulations," The Journal of Chemical Physics, vol. 157, no. 11, Sep. 2022. doi:10.1063/5.0106617
70. P. Hirel, "Atomsk: A tool for manipulating and converting Atomic Data files," Computer Physics Communications, vol. 197, pp. 212–219, Dec. 2015. doi:10.1016/j.cpc.2015.07.012
71. A. Stukowski, "Visualization and analysis of atomistic simulation data with OVITO–The Open Visualization Tool," Modelling and Simulation in Materials Science and Engineering, vol. 18, no. 1, p. 015012, Dec. 2009. doi:10.1088/0965-0393/18/1/015012
72. T. Beuerle, K. Hummler, C. Elsässer, and M. Fähnle, "Formation of magnetic moments on iron impurities in transition-metal hosts," Physical Review B, vol. 49, no. 13, pp. 8802–8810, Apr. 1994. doi:10.1103/physrevb.49.8802
73. J. Zhang, D. J. M. King, P. A. Burr, A. P. Horsfield, and M. R. Wenman, "The importance of many body interactions in point defect clusters in α zirconium alloys," Journal of Nuclear Materials, vol. 620, p. 156308, Jan. 2026. doi:10.1016/j.jnucmat.2025.156308
74. Zhao, Y., Wang, C., Peng, Q., & Lu, G. "Error analysis and applications of a general QM/MM approach." Computational Materials Science, 50(2), 714–719., Dec. 2010. doi:10.1016/j.commatsci.2010.09.038
75. Barrett, R., Dietschreit, J. C. B., & Westermayr, J. "Incorporating long-range interactions via the multipole expansion into ground and excited-state molecular simulations." npj Computational Materials, 12, 135, Mar. 2026. doi:10.1038/s41524-026-02048-3.
76. Birks, F., Nutter, M., Swinburne, T. D., et al. "Efficient and accurate spatial mixing of machine learned interatomic potentials for materials science." npj Computational Materials, 12, 110, Feb. 2026. doi:10.1038/s41524-026-01982-6.

## Supplementary Information

Elastic constant predictions from the universal MACE and fine-tuned NEP potentials for Zr and Fe, compared with DFT.

| Method | Material | C11 GPa | C12 GPa | C13 GPa | C33 GPa | C44 GPa | C66 GPa |
|---|---|---|---|---|---|---|---|
| **DFT** | **Zr** | **141.1** | **67.6** | **64.3** | **166.9** | **25.8** | **36.8** |
| MACE | Zr | 144.3 | 65.8 | 69.0 | 160.8 | 35.8 | 39.2 |
| **DFT** | **Fe** | **271** | **145** | **--------** | **--------** | **101** | **--------** |
| MACE | Fe | 295.5 | 162.6 | -------- | -------- | 130.2 | -------- |
| NEP-ft | Fe | 275.1 | 94.6 | -------- | -------- | 112.3 | -------- |

Defect formation energy predictions from the universal MACE and fine-tuned NEP potentials for Zr and Fe, compared with DFT. With s for substitutional solutes and i for interstitial solutes.

| Zr system | Vacancy eV | Sn-s eV | Nb-s eV | Fe-i eV | H-i eV |
|---|---|---|---|---|---|
| DFT | 2.06 | -1.16 | 0.60 | 0.93 | -0.45 |
| MACE | 2.25 | -1.17 | 0.52 | 1.01 | -0.55 |

| Fe system | Vacancy eV | Si-s eV | C-i eV |
|---|---|---|---|
| DFT | 2.19 | -1.21 | 0.72 |
| MACE | 2.20 | -1.13 | 1.05 |
| NEP-ft | 2.23 | -1.28 | 1.54 |

The NEP potential fine-tuning training parameters used. $\lambda 1$ is the L1 regularisation parameter, $\lambda 2$ is the L2 regularisation parameter, $\lambda e$ is the relative energy training weight, $\lambda f$ is the relative forces training weight, and $\lambda v$ is the relative virials training weight. Nbatch is the batch size, Npopulation is the population size, Ngenerations is the number of generations and Ndataset is the total fine-tuning dataset size used for the separable natural evolutionary strategy (SNES) used by NEP.

| Parameter | Value |
|---|---|
| $\lambda 1$ | 0 |
| $\lambda 2$ | 0.1 |
| $\lambda e$ | 1 |
| $\lambda f$ | 1 |
| $\lambda v$ | 1 |
| Nbatch | 49 |
| Npopulation | 50 |
| Ngenerations | 500 |
| Ndataset | 49 |

Definitions of the energy terms used in the QM/ML framework for benchmarking. QM/ML denotes energies obtained using DFT–MLIP mixing for the full four-region model. QM/ML–no outer buffer refers to the same mixing applied only to the inner three regions, with the fixed outer-buffer atoms excluded. QM corresponds to the DFT energies of the QM plus inner-buffer region. Energy terms with the suffix "–MLIP" are calculated using the MLIP alone, without DFT–MLIP mixing.

| Energy type | Reference energy | Region contained |
|---|---|---|
| QM/ML | DFT | All regions |
| QM/ML-MLIP | MLIP | All regions |
| QM/ML-no outer buffer | DFT | QM+inner buffer+ML |
| QM/ML -no outer buffer-MLIP | MLIP | QM+inner buffer+ML |
| QM-DFT | DFT | QM+inner buffer |
| QM-MLIP | MLIP | QM+inner buffer |

The maximum, minimum and mean absolute force differences were evaluated relative to the 10 k-point reference forces. Assuming a force-convergence criterion of 0.1 eV $Å^{-1}$, k-point meshes of four or more were sufficient for the QM region alone, while meshes of at least six were required for the QM + inner buffer region containing the vacuum interface.

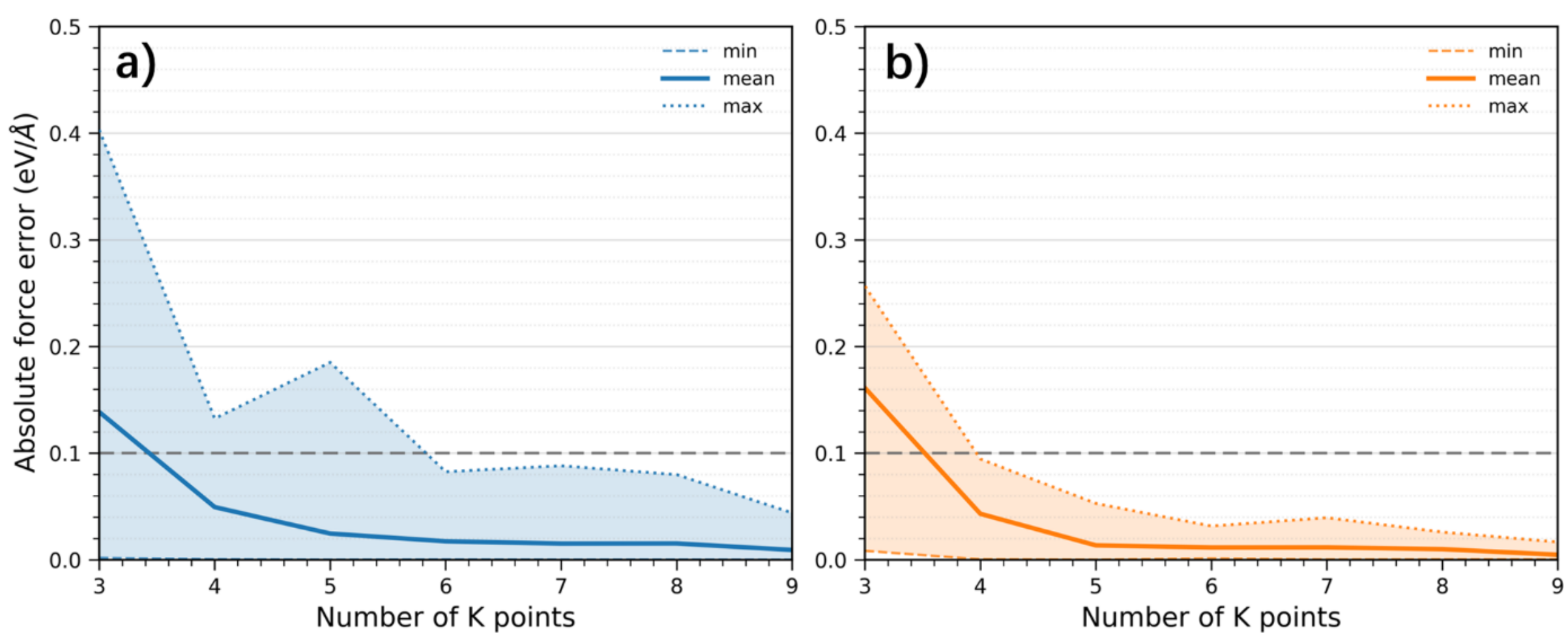

*Absolute DFT force error relative to the 10 K points reference. (a) QM + inner buffer region, (b) QM region.*

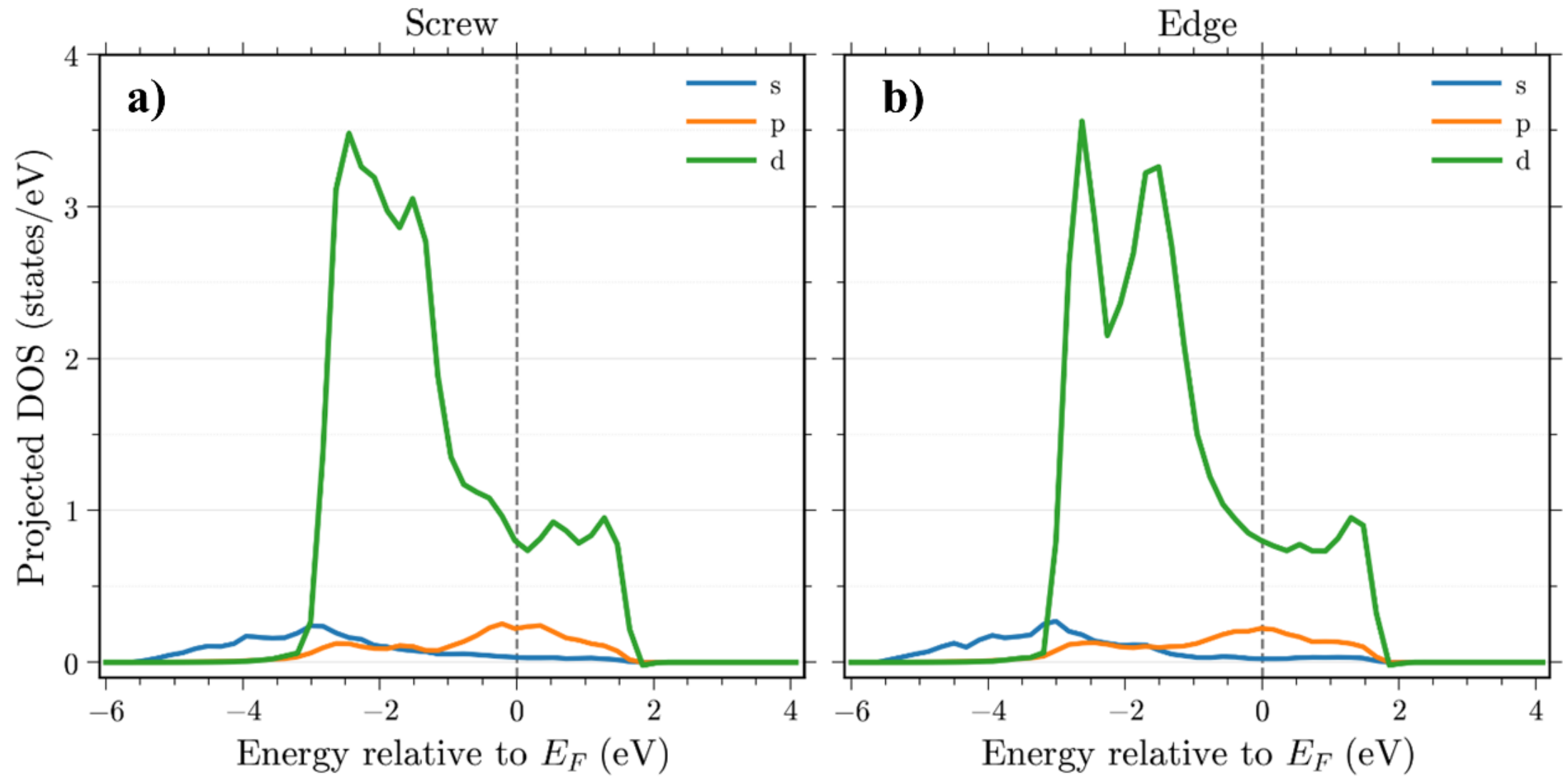

*Orbital-resolved PDOS for Fe interstitial solutes bound to (a) ⟨a⟩ screw and (b) ⟨c/2⟩ edge dislocations in Zr. Contributions from Fe s, p, and d orbitals are shown, with energies referenced to the Fermi level. Only Fe states are included. The PDOS is used to assess the validity of the non-spin-polarised DFT treatment.*

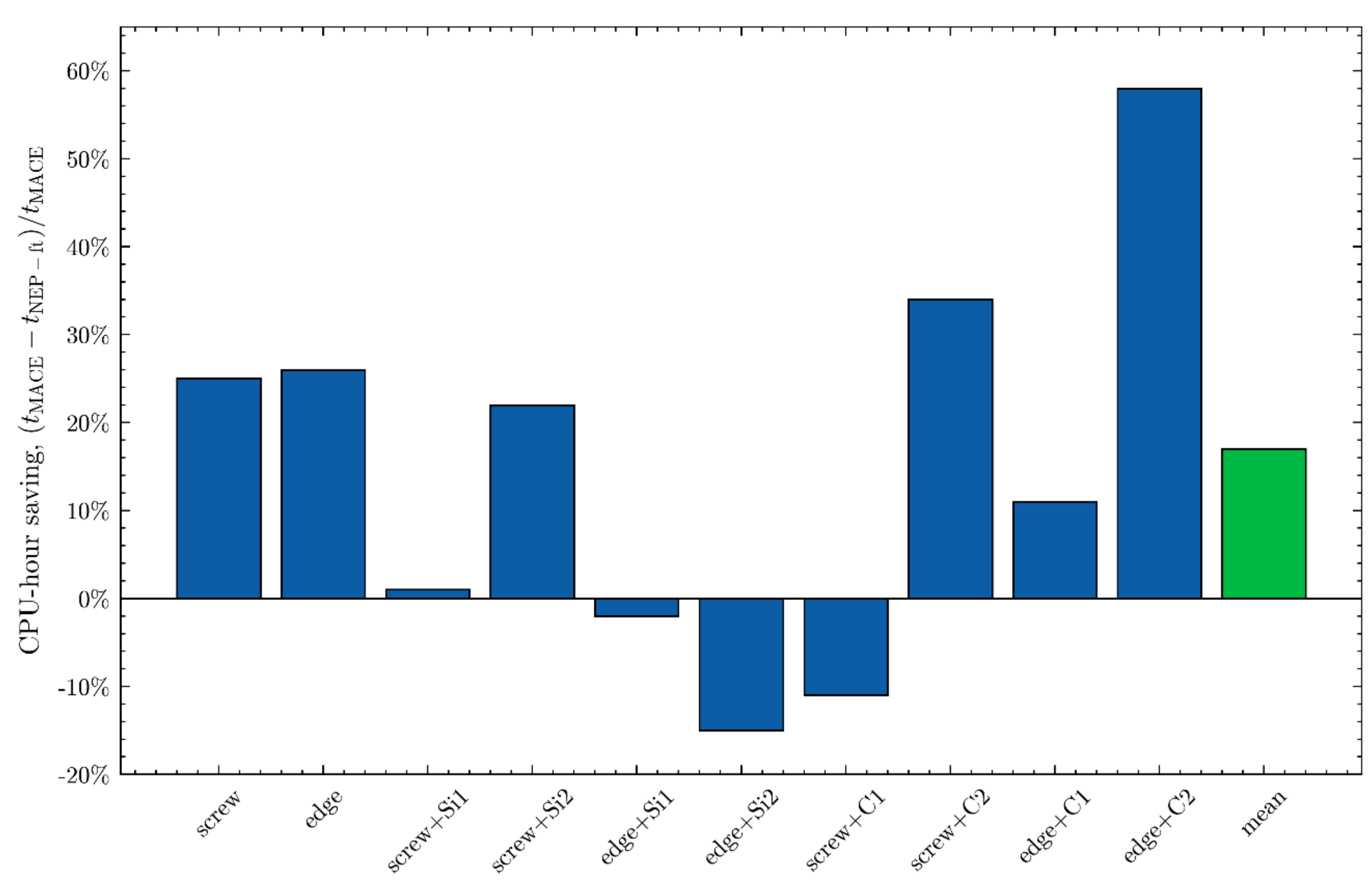

*Relative CPU-hour savings for QMML simulations using the NEP-ft potential instead of the MACE potential in BCC Fe systems with and without solutes. Positive values indicate that NEP-ft requires fewer CPU hours than MACE for the similar simulation, along with the mean across all systems is shown.*